\documentclass[11pt,a4paper]{article}

\usepackage{amsmath,amsfonts,graphicx,natbib,psfrag,color}
\usepackage{algorithm, algorithmic}
\usepackage[textsize=small]{todonotes}
\usepackage{booktabs}
\usepackage{cleveref}
\usepackage[parfill]{parskip}
\usepackage{multirow}
\usepackage{url}

\usepackage{tikz}
\usetikzlibrary{shapes.geometric, arrows, positioning, arrows.meta, shapes.misc}
\tikzstyle{b} = [rectangle, rounded corners, minimum width=2cm, minimum height=2cm,text centered, draw=black, fill=blue!30]
\tikzstyle{arrow} = [thick,->,>={Stealth[scale=1.5]}]

\title{Accelerating Bayesian inference for stochastic epidemic models using incidence data}

\author{Andrew Golightly$^1$\footnote{andrew.golightly@durham.ac.uk}, Laura E. Wadkin$^2$, Sam A. Whitaker$^2$, Andrew W. Baggaley$^2$,\\ Nick G. Parker$^2$, Theodore Kypraios$^3$}
\date{\small $^1$ Department of Mathematical Sciences, Durham University, Stockton Road, DH1 3LE, UK\\$^2$ School of Mathematics, Statistics and Physics, Newcastle University, NE1 7RU, UK\\$^3$ School of Mathematical Sciences, University Of Nottingham, NG7 2RD, UK}

\begin{document}
\maketitle
\begin{abstract}
\noindent We consider the case of performing Bayesian inference for stochastic epidemic compartment models, using incomplete time course data consisting of incidence counts that are either the number of new infections or removals in time intervals of fixed length. We eschew the most natural Markov jump process representation for reasons of computational efficiency, and focus on a stochastic differential equation representation. This is further approximated to give a tractable Gaussian process, that is, the linear noise approximation (LNA). Unless the observation model linking the LNA to data is both linear and Gaussian, the observed data likelihood remains intractable. It is in this setting that we consider two approaches for marginalising over the latent process: a correlated pseudo-marginal method and analytic marginalisation via a Gaussian approximation of the observation model. We compare and contrast these approaches using synthetic data before applying the best performing method to real data consisting of removal incidence of oak processionary moth nests in Richmond Park,  London. Our approach further allows comparison between various competing compartment models.     
\end{abstract}


\section{Introduction}
\label{sec:intro}

The starting point for formulating a model of epidemic spread is usually a set of compartments or classes, characterising the individuals participating in the epidemic \citep[see e.g.][]{jacquez72}. A transmission model then describes the dynamics of individuals within each class, and this can be combined with models of severity and detection/observation \citep[see e.g.][]{birrell20} to give an overarching model that links epidemic data to latent disease transmission. The transmission model can be deterministic or stochastic, with the former typically taking the form of a coupled, nonlinear ordinary differential equation (ODE) system and the latter a continuous-time Markov jump process (MJP). The link between deterministic and stochastic models is made clear in \cite{kurtz1970} \citep[see also][]{kurtz1971}, with the ODE approach seen as a large population limit of the MJP. However, while fitting ODE models in the presence of a simple observation model is relatively straightforward, they ignore intrinsic stochasticity, which may play an important role in the dynamics of the epidemic, particularly at the start or the end of an outbreak, when numbers of individuals with the disease are likely to be small. Stochastic transmission models have been widely adopted for small-size epidemics \citep[see e.g.][]{boys2007,STOCKDALE19,FUNK18} whereas \cite{corbella22} \citep[see also][]{birell2011} combine a deterministic transmission model with sophisticated stochastic observation models. 

Although complex stochastic epidemic models can provide an accurate description of disease transmission, fitting such models to discrete-time data that may be incomplete and subject to error is typically complicated by the intractability of the observed data likelihood. Earlier attempts to address this issue include the use of data augmentation and Metropolis-Hastings steps to integrate over the uncertainty associated with unobserved dynamic components \citep{gibson1998,oneill1999}. Recent attention has focused on approaches that only require the ability to generate forward realisations from the transmission model such as approximate Bayesian computation \citep{mckinley2009,KYPRAIOS17} and pseudo-marginal methods \citep{mckinley2014,spannaus20,corbella22}. Nevertheless, these methods typically require many (millions of) model simulations which can be impracticable for large-size epidemics, that is, epidemics in which the population size is greater than a few thousand individuals. Alternative approaches that aim to reduce the computational burden include the use of cheap but approximate ``surrogate'' methods including the use of Gaussian process emulation \citep{Scarponi2022,Swallow22}, direct approximation of the MJP e.g. via a linear noise approximation \citep[LNA,][]{fintzi2021linear} and tractable approximation of the observed data likelihood \citep[see e.g.][]{whiteley21}.                   

In this paper, we consider incidence data consisting of the number of new cases (either infections or removals) accumulated in an observation interval. We further assume imperfect observations and consider two approaches commonly used to allow for under reporting or over dispersion. We build on the work of \cite{fintzi2021linear} by adopting a linear noise approximation of the latent cumulative incidence process. Whereas 
\cite{fintzi2021linear} integrate over the uncertainty in the latent process via a sampling approach, we introduce a further Gaussian approximation of the observation model, allowing analytic integration of the latent incidence process. Our framework additionally allows for a time-varying infection rate, as is typically required for accounting for seasonality and/or interventions. The infection rate is modelled stochastically as an additional component in the system of stochastic differential equations which the LNA approximates. Hence, our contribution is a fast and efficient sampling-based framework for inferring the parameters of a general class of stochastic epidemic models. We benchmark the performance of our approach against state-of-the-art correlated pseudo-marginal methods \citep{dahlin2015,deligiannidis2018} in terms of both accuracy and efficiency, using a susceptible-infectious-removed (SIR) model. 

Finally, we consider the application of the methodology to the infestation of the oak processionary moth (OPM),  \textit{Thaumetopoea processionea}, in Richmond Park, London. Using time course data consisting of the yearly removal incidence of infested trees between the years 2013 and 2020, we compare and contrast an SIR model with a model in which infected trees can re-enter the susceptible class. The assumed initial population size is some 40,000 oak trees with the number of susceptible trees reducing to around 35,000 over the time frame of the data set. The size of the epidemic necessitates analytic integration of the latent incidence process; discrete stochastic models combined with exact (simulation-based) inference methods such as data augmentation \citep[see e.g.][]{jewell09} are practically infeasible here; see \cite{stockdale21} for a recent discussion.  

The remainder of this paper is organised as follows. Stochastic epidemic models, and in particular, the jump process and subsequent LNA representations of the cumulative incidence process are considered in Section~\ref{sec:sem}. The inference task is described in Section~\ref{sec:binf}, including marginalisation of the incidence process via pseudo-marginal and analytic methods, with the latter using an additional approximation. Applications are considered in Section~\ref{sec:app} with a discussion of limitations of the proposed approach in Section~\ref{sec:sum}.

\section{Stochastic epidemic models}
\label{sec:sem}
For ease of exposition, we consider an SIR epidemic model \citep{AnBr00,kermack27} within which a population of fixed size $N_{\textrm{pop}}$ is classified into compartments consisting of susceptible ($S$), infectious ($I$) and removed ($R$) individuals. However, we note that the framework can be easily extended to allow for additional compartments e.g. SEIR models \citep{hethcote2000} which allows for susceptibles entering an exposed class prior to becoming infectious. 

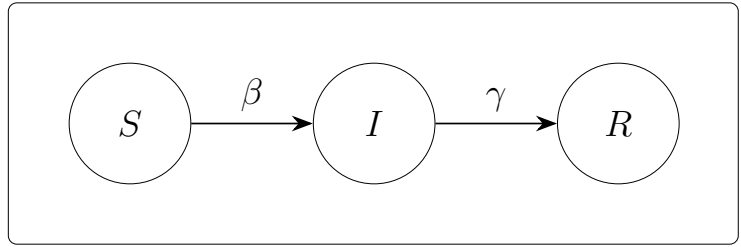
\begin{figure}[h]
\centering
\scalebox{0.8}{
{\LARGE
\begin{tikzpicture}
\draw [rounded corners,fill=white!20] (-2,-2) rectangle (10,2);
\node (s) [draw, circle, minimum height=2cm,fill=white] {$S$};
\node (i) [draw, circle, xshift=1cm, minimum height=2cm,right = of s,fill=white] {$I$};
\node (r) [draw, circle, xshift=1cm, minimum height=2cm,right = of i,fill=white] {$R$};
\draw [arrow] (s) -- node[anchor=south] {${\beta}$} (i);
\draw [arrow] (i) -- node[anchor=south] {$\gamma$} (r);
\end{tikzpicture}
}
}
\caption{SIR compartment model.}\label{fig:SIR}
\end{figure}
The SIR compartment model is shown in Figure~\ref{fig:SIR}. Transitions between compartments can be described by the set of pseudo-reactions given by
\[
S+I \xrightarrow{\beta} 2I, \quad I \xrightarrow{\gamma} R . 
\]
The first transition describes contact of an infective individual with a susceptible and with the net effect resulting in an additional infective individual and one fewer susceptible. The final transition accounts for removal (recovered with immunity, quarantined or dead) of an infected individual. The components of $\theta=(\beta,\gamma)'$ denote infection and removal rates.

In what follows, we describe the most natural stochastic SIR model as a Markov jump process before considering a linear noise approximation suitable for incidence data.  

\subsection{Markov jump process}
\label{sec:mjp}
Let $X_t=(S_t,I_t)'$ denote the numbers in each state at time $t\geq 0$ and note that $R_t=N_{\textrm{pop}}-S_t-I_t$ for all $t\geq 0$. We denote by $\theta=(\beta,\gamma)'$ the vector of parameters. The dynamics of $\{X_t,t\geq 0\}$ are most naturally described by a Markov jump process (MJP), that is, a continuous time, discrete valued Markov process. Assuming that at most one event can occur over a time interval $(t,t+\Delta t]$ and that the state of the system at time $t$ is $x_t=(s_t,i_t)'$, the MJP representation of the prevalence process is characterised by transition probabilities of the form 
\begin{align*}
\mathbb{P}(X_{t+\Delta t}=(s_t-1,i_t+1)'|x_t,\theta) &= \beta s_t i_t \,\Delta t+o(\Delta t),\\
\mathbb{P}(X_{t+\Delta t}=(s_t,i_t-1)'|x_t,\theta) &= \gamma i_t \,\Delta t+o(\Delta t),\\
\mathbb{P}(X_{t+\Delta t}=(s_t,i_t)'|x_t,\theta) &= 1-(\beta s_t i_t + \gamma i_t)\,\Delta t+o(\Delta t),
\end{align*}
where $o(\Delta t)/\Delta t\to 0$ as $\Delta t\to 0$. Similarly, the cumulative incidence of infection and removal events $\{N_t, t\geq 0\}$ is an MJP governed by the transition probabilities 
\begin{align*}
\mathbb{P}(N_{t+\Delta t}=(n_{1,t}+1,n_{2,t})'|n_t,x_t,\theta) &= \beta s_t i_t \,\Delta t+o(\Delta t),\\
\mathbb{P}(N_{t+\Delta t}=(n_{1,t},n_{2,t}+1)'|n_t,x_t,\theta) &= \gamma i_t \,\Delta t+o(\Delta t),\\
\mathbb{P}(N_{t+\Delta t}=(n_{1,t},n_{2,t})'|n_t,x_t,\theta) &= 1-(\beta s_t i_t + \gamma i_t)\,\Delta t+o(\Delta t),
\end{align*}
where $n_t=(n_{1,t},n_{2,t})'$ denotes the cumulative number of infection and removal events at time $t$. 

The processes describing the prevalence $\{X_t,t\geq 0\}$ and cumulative incidence $\{N_t, t\geq 0\}$ are linked via the equation
\begin{equation}\label{MJP1}
X_t = x_0 + \sum_{i=1}^2 S^i N_{i,t}
\end{equation}
where $x_0$ represents the initial number of susceptible and infected individuals, $N_{i,t}$ denotes the $i$th component of the incidence process at time $t$ and $S^i$ is the $i$th column of the stoichiometry matrix
\[
S=\begin{pmatrix}
 -1 & 0\\
  1 & -1
\end{pmatrix}.
\]
Given $x_0$ and $\theta$, generating exact realisations of the incidence process, and therefore the stochastic SEIR model is straightforward and can be achieved by using well known simulation algorithms from the stochastic kinetic models literature \citep[see e.g.][]{wilkinson2018}. At this point, it will be helpful to define the instantaneous rate or hazard function $h(x_t)=(h_1(x_t),h_2(x_t))'$ by 
\begin{align*}
h(x_t)&=\lim_{\Delta t\to 0} \mathbb{P}(X_{t+\Delta t}|x_t,\theta) / \Delta t\\
&= (\beta s_t i_t, \gamma i_t)'
\end{align*}
for $X_{t+\Delta t}$ resulting from either an infection or removal reaction respectively. Note that $\theta$ is suppressed here for notational convenience. Then, given a state $x_t$ at time $t$, Gillespie's direct method \citep{gillespie1977exact} simulates the time to the next event as an exponential random variable with rate $\sum_{i=1}^2 h_i$. The event that occurs will be of type $i$ (with $i=1$ infection, $i=2$ removal) with probability proportional to $h_i$. 

\subsection{Linear noise approximation (LNA)}
\label{sec:lna}

The linear noise approximation (LNA) is most commonly presented as a Gaussian process approximation of the MJP description of the prevalence process $\{X_t, t\geq 0\}$ (see e.g. \cite{ross09,fearnhead14,Fuchs_2013} in the epidemic context and \cite{ferm2008,Komorowski09,stathopoulos13} in a wider systems biology context). As in \cite{fintzi2021linear}, we require an approximation of the cumulative incidence process $\{N_t,t\geq 0\}$. Although this can be derived in a number of more or less formal ways, we give an intuitive derivation along the lines of \cite{wallace10}.

Consider an infinitesimal time interval, $(t,t+dt]$, over which the
reaction hazards will remain constant almost surely. Consequently, 
the counting process over this interval for the $i$th 
component, denoted by $dN_{i,t}$, is Poisson distributed 
with rate $h_i dt$. Stacking these quantities in the vector $dN_t$
 it should be clear that
\[
\operatorname{E}(dN_t)=h(x_t)dt,\qquad \operatorname{Var}(dN_t)= \operatorname{diag}\{h(x_t)\}dt,
\]
Hence, the It\^o Stochastic differential equation (SDE) 
that best matches the MJP representation of the incidence process is given by 
\begin{equation}\label{cle}
dN_t = h(x_t)dt + \operatorname{diag}\{\sqrt{h(x_t)}\}\,dW_t,
\end{equation}
where $W_t$ is a length-2 vector of uncorrelated standard Brownian motion processes and $\operatorname{diag}\{\sqrt{h(x_t)}\}$ 
is a $2\times 2$ diagonal matrix with non-zero entries given by $\sqrt{h_1(x_t)}$ and $\sqrt{h_2(x_t)}$. Note that the RHS of (\ref{cle}) can be written explicitly in terms of $n_t$ via (\ref{MJP1}) which gives $x_t=(s_0-n_{t,1},i_0+n_{t,1}-n_{t,2})'$. We then further define the hazard function written in terms of $n_t$ as 
\[
h^*(n_t)=(\beta [s_0-n_{t,1}][i_0+n_{t,1}-n_{t,2}],\gamma [i_0+n_{t,1}-n_{t,2}])'
\] 
for which (\ref{cle}) becomes
\begin{equation}\label{cle2}
dN_t = h^*(n_t)dt + \operatorname{diag}\{\sqrt{h^*(n_t)}\}\,dW_t.
\end{equation}

The SDE in (\ref{cle2}) can be linearised as follows. Consider a partition of $N_t$ as $N_t=\eta_t +R_t$ where 
$\eta_t$ is a deterministic process satisfying the ordinary differential equation (ODE)
\begin{equation}\label{lna1}
\frac{d\eta_t}{dt} = h^*(\eta_t)
\end{equation} 
and $R_t=N_t-\eta_t$ is a residual stochastic process satisfying the (typically) intractable SDE
\begin{equation}\label{eqn:resid_sde}
dR_t=\{h^*(n_t)-h^*(\eta_t)\}\,dt+\operatorname{diag}\{\sqrt{h^*(n_t)}\}\,dW_t.
\end{equation}
We obtain an approximate, tractable $\hat{R}_t$ by Taylor expanding $h^*(n_t)$ 
and $\operatorname{diag}\{h^*(n_t)\}$ about $\eta_t$. Retaining the first two terms in the 
expansion of the former and the first term in the expansion of the latter gives
\begin{equation}\label{eqn:approx_resid_sde}
d\hat{R}_t=F_t\hat{R}_t\,dt+ \operatorname{diag}\{\sqrt{h^*(\eta_t)}\}\,dW_t
\end{equation}
where $F_t$ is the Jacobian matrix with ($i$,$j$)th element given by the partial derivative of the $i$th component of $h^*(\eta_t)$ with respect to the $j$th component of $\eta_t$. Hence, we obtain
\[
F_t=\begin{pmatrix}
\beta (s_0-i_0-2\eta_{t,1}+\eta_{t,2}) & \beta (\eta_{t,1}-s_0) \\
\gamma & -\gamma
\end{pmatrix}.
\]  
The solution of (\ref{eqn:approx_resid_sde}) is straightforward to obtain \citep[see e.g.][among several others]{fearnhead14}. Omitting details, we arrive at
\begin{equation}\label{lnaINC}
N_t | N_{0}=\left(\eta_0+r_0 \right) \sim \textrm{N}(\eta_t + G_t r_{0}, V_t)
\end{equation}
where the fundamental matrix $G_t$ satisfies
\begin{equation} \label{lna2}
\frac{dG_t}{dt} = F_t G_t, \quad G_{0} = I_2
\end{equation}
and $V_t$ satisfies
\begin{equation} \label{lna3}
\frac{dV_t}{dt} = V_t F_t' + \operatorname{diag}\{h^*(\eta_t)\} + F_t V_t, \quad V_0 = 0_{2}.
\end{equation}
Note that $I_2$ and $0_2$ are the $2\times 2$ identity and zero matrices respectively. The LNA for the cumulative 
incidence process is then summarised by (\ref{lnaINC}) and (\ref{lna1}), (\ref{lna2}), (\ref{lna3}).

\subsection{Time varying infection rate}
In practice it may be unreasonable to assume that the infection rate in the SIR model remains constant throughout the epidemic (e.g. due to seasonality and/or interventions). We therefore follow \cite{dureau13} and \cite{spannaus20} (among others) and describe the contact rate via an It\^o stochastic differential equation (SDE). Let $\{\beta_t,t\geq 0\}$ denote the infection process and consider $N_{3,t} = \log(\beta_t)$, assumed to satisfy a time-homogeneous SDE of the form
\begin{equation}
        {d}N_{3,t} = a(n_{3,t})
        {d}t + 
        b(n_{3,t}){d}W_{3,t}
        \label{eqn:TVB}
\end{equation}
where $\{W_{3,t},t \geq 0\}$ is a standard Brownian motion process. Combining (\ref{eqn:TVB}) with (\ref{cle2}) gives a coupled SDE for $N_t=(N_{1,t},N_{2,t},N_{3,t})'$ of the form
\begin{equation} \label{cle3}
dN_{t} = \left\{h_{1}^*(n_t),h_{2}^*(n_t),a(n_{3,t}) \right\} dt + \operatorname{diag}\left\{\sqrt{h^*_1(n_t)},\sqrt{h^*_2(n_t)},b(n_{3,t})\right\}\, dW_t
\end{equation}
where $W_t$ is a length-3 vector of uncorrelated standard Brownian motion processes. The LNA of (\ref{cle3}) follows in the same way as Section~\ref{sec:lna}, albeit with the Jacobian matrix $F_t$ redefined as 
\[
F_t=\begin{pmatrix}
\exp(\eta_{3,t}) (s_0-i_0-2\eta_{t,1}+\eta_{t,2}) & \exp(\eta_{3,t}) (\eta_{t,1}-s_0) & \exp(\eta_{3,t})(s_0-\eta_{t,1})(i_0+\eta_{t,1}-\eta_{t,2}) \\
\gamma & -\gamma & 0 \\
0 & 0 & \frac{\partial a (\eta_{3,t})}{\partial \eta_{3,t}}
\end{pmatrix}
\]  
and the RHS of (\ref{lna1}) and (\ref{lna3}) augmented to include $a(n_{3,t})$ and $b^2(n_{3,t})$.

\section{Bayesian inference}
\label{sec:binf}
In this section, we consider the problem of performing fully Bayesian inference for the parameters (and unobserved dynamic processes) governing the SIR (LNA) model, based on incidence observations that we assume are incomplete and subject to measurement error. We describe the observation model before considering the inference task. For ease of exposition, we assume a constant infection rate, but note that extension of the methodology to a time varying infection rate is straightforward and considered in Section~\ref{sec:app}.  

\subsection{Observation model}
\label{sec:obsmodel}
Without loss of generality, consider data $y=(y_{1},\ldots, y_{T})'$ at integer times, where $y_{t}$ is a (partial) observation on the cumulative incidence $\Delta N_{t}=N_{t}-N_{t-1}$ over a time interval $(t-1,t]$. Commonly used models for incidence data include additive Gaussian noise \citep{dureau13}, the Binomial distribution \citep{cauchemez08} and the Negative Binomial distribution \citep{LloydSmith07,fintzi2021linear,spannaus20}. The latter two models are typically used under the assumption of under reporting and to capture overdispersion, respectively. In large population settings, they may be well approximated by a Gaussian distribution which may offer computational benefits when combined with a Gaussian description of the underlying epidemic dynamics, such as the LNA described in Section~\ref{sec:lna}. These models take the form
\begin{align}
	Y_t | \left(\Delta N_t=\Delta n_t\right) &\sim \textrm{N}\left(P'\Delta n_{t}, \sigma^2 \right),\label{obsNorm}\\
	Y_{t}|\left(\Delta N_{t}=\Delta n_t\right) &\sim \textrm{Bin}\left(P'\Delta n_{t},\lambda\right),\label{obsBin}\\
        Y_{t}|\left(\Delta N_{t}=\Delta n_t\right) &\sim \textrm{NegBin}\left(\mu=\lambda P'\Delta n_{t},\sigma^2=\mu+\phi \mu^2\right)\label{obsNBin}
\end{align}
for $t=1,\ldots,T$. We assume that $Y_t$ is univariate and $P$ is a constant matrix with $P'=(1,0)$ corresponding to noisy counts of new infections in a given time window and $P'=(0,1)$ for removals. We further assume that the observations are independent (given the latent process) and we let $\pi(y_{t}|\Delta n_{t},\psi)$ denote the probability mass function linking $y_t$ and $\Delta n_t=n_t-n_{t-1}$, with $\psi$ denoting the parameters governing the observation model. For example, in (\ref{obsNBin}), $\psi=(\lambda,\phi)'$ with $\lambda$ controlling the average proportion of cases seen and $\phi$ is the (inverse) overdispersion parameter. 

\subsection{Inference task}
\label{sec:inftask}
We assume that interest lies in the vector of all static parameters $\theta=(\beta,\gamma,\psi',x_0')'$ including the initial state $x_0=(s_0,i_0)'$ for convenience, and the latent incidence process at the observation times $n=\{n_t, t=0,\ldots,T\}$.  

Upon ascribing a prior density $\pi(\theta)$ to $\theta$, Bayesian inference proceeds via the joint posterior
\begin{align}\label{jpost}
\pi(\theta, n|y)& \propto \pi(\theta) \pi(n|x_0,\beta,\gamma)\pi(y|\Delta n,\psi)
\end{align}
where
\[
\pi(n|x_0,\beta,\gamma) = \prod_{t=1}^{T} \pi(n_{t}| n_{t-1},x_0,\beta,\gamma)
\]
and $\pi(n_{t}| n_{t-1},x_0,\beta,\gamma)$ is the Gaussian transition density obtained from (\ref{lnaINC}) with the ODEs (\ref{lna1}) and (\ref{lna3}) integrated over $[t-1,t]$ with $\eta_{t-1}=n_{t-1}$ and $V_{t-1}=0_{2}$. Note that in this case the residual 
$r_{t-1}=n_{t-1}-\eta_{t-1}=0$ and subsequently the ODE satisfied by $G_t$ in (\ref{lna2}) need not be integrated. Finally, 
\[
\pi(y|\Delta n,\psi) = \prod_{t=1}^{T} \pi(y_{t}|\Delta n_{t},\psi).
\]
Since the joint posterior in (\ref{jpost}) will be intractable, we resort to Monte Carlo methods for generating samples of the parameters and latent dynamic process. A Gibbs sampler provides a natural mechanism for sampling (\ref{jpost}), whereby one alternates between draws of $\theta|n,y$ and $n|\theta,y$. However, dependence between $n$ and $\theta$ can lead to poor mixing. For this reason, \cite{fintzi2021linear} use a non-centred parameterisation whereby standard Gaussian innovations driving the generative form of the LNA are used as the effective components to be conditioned on. In what follows, we take a different approach by marginalising out the latent process, either via (correlated) pseudo-marginal methods \citep[e.g.][]{andrieu10,deligiannidis2018} or by further approximating the observation model in the non Gaussian case.

\subsection{Marginalisation of the incidence process}
\label{sec:marg}

The joint posterior density in (\ref{jpost}) can be factorised as 
\begin{equation}\label{jpost2}
\pi(\theta, n|y) = \pi(\theta|y)\pi(n|\theta,y)
\end{equation}
where
\begin{equation}\label{jpost3}
\pi(\theta|y) \propto \pi(\theta)\pi(y|\theta).
\end{equation}
The form of (\ref{jpost2}) suggests a two step approach to inference whereby samples are first drawn from the marginal parameter posterior $\pi(\theta|y)$ in step 1, and then conditioned on in a second step when drawing samples of the latent process from $\pi(n|\theta,y)$. However, unless the observation model takes the linear Gaussian form of (\ref{obsNorm}), neither the observed data likelihood $\pi(y|\theta)$ in (\ref{jpost3}) nor the constituent densities in (\ref{jpost2}) will be tractable, despite the linear Gaussian structure of the LNA. The main focus of this paper is exactly this \emph{intractable} scenario, and we now consider two approaches to address it.   

\subsubsection{Pseudo-marginal methods} 
\label{sec:pmmh}

Consider first the intractable observed data likelihood $\pi(y|\theta)$ which can be factorised as 
\begin{equation}\label{marglike}
\pi(y|\theta)=\pi(y_1|\theta)\prod_{t=2}^T \pi(y_t|y_{1:t-1},\theta)
\end{equation}
where $y_{1:t-1}=(y_1,\ldots,y_{t-1})'$. The terms in (\ref{marglike}) can be recursively estimated using a particle filter \citep[see e.g.][for an overview]{chopin2020introduction} in such a way that realisations of a non-negative unbiased estimator of the full likelihood are obtained. We denote this estimator by 
\[
\hat{\pi}_{U}(y|\theta)=\hat{\pi}_{U_{1}}(y_1|\theta)\prod_{t=2}^T \hat{\pi}_{U_{t}}(y_t|y_{1:t-1},\theta)
\] 
where the flattened vector $U=(U_1',\ldots,U_T')'\sim g(u)$ denotes all random variables used in the construction of the estimator. Hence, unbiasedness here means that $E_{U\sim g}\{\hat{\pi}_{U}(y|\theta)\}= \pi(y|\theta)$. Algorithm~\ref{auxPF} gives step $t+1$ of the particle filter and can be executed for $t=0,\ldots,T-1$ upon initialising with particles $\{n_{0}^{(k)}=0_2, k=1,\ldots,N\}$. Note that component $t+1$ of $U$ is partitioned as $(\tilde{U}_{t+1}',\bar{U}_{t+1})'$ where the $k$th value of $\tilde{U}_{t+1}$ is $\tilde{U}_{k,t+1}\sim \textrm{N}(0,I_2)$ and used to propagate particles in step 1(b); $\bar{U}_{t+1}\sim \textrm{Unif}(0,1)$ is used in the systematic resampling step 3. As presented, step 1a has that the ODE system governing the LNA is initialised and integrated for each particle $n_t^{(k)}$. This `restarting' approach avoids potential mismatch between the deterministic process $\eta_t$ and the latent stochastic process $n_t$ \citep[see e.g.][for further discussion]{fearnhead14,minas17}. Upon iterating Algorithm~\ref{auxPF} over $t$, the product (over observation times) of the average unnormalised weight gives an unbiased estimator of $\pi(y|\theta)$ \citep{delmoral04} and is key to the construction of a pseudo-marginal scheme that we now describe. 

\begin{algorithm}[t]
\caption{Step $t+1$ of the Particle Filter}\label{auxPF}
Input: Parameter $\theta$, auxiliary variable $u_{t+1}=(\tilde{u}_{t+1}',\bar{u}_{t+1})'$, next observation $y_{t+1}$, $N$ particles $\{n_{t}^{(k)}, k=1,\ldots,N\}$.

\begin{enumerate}
\item Forward propagation. For $k=1,\ldots,N$: 
\begin{itemize}
\item[(a)] Integrate (\ref{lna1}) and (\ref{lna3}) over $(t,t+1]$ with initial conditions $\eta_t^{(k)}=n_t^{(k)}$ and $V_t^{(k)}=0_2$ to give $\eta_{t+1}^{(k)}$ and $V_{t+1}^{(k)}$. Note that (\ref{lna2}) need not be integrated.
\item[(b)] Set $n_{t+1}^{(k)}=\eta_{t+1}^{(k)}+\sqrt{V_{t+1}^{(k)}}\, \tilde{u}_{k,t+1}$ and 
$\Delta n_{t+1}^{(k)}=n_{t+1}^{(k)}-n_{t}^{(k)}$.
\end{itemize}
\item Compute the weights. For $k=1,\ldots,N$:
\[
\tilde{w}_{t+1}^{(k)}=\pi\left(y_{t+1}|\Delta n_{t+1}^{(k)},\psi\right), \qquad w_{t+1}^{(k)}=\frac{\tilde{w}_{t+1}^{(k)}}{\sum_{j=1}^{N}\tilde{w}_{t+1}^{(j)}}.
\]
\item Resample $N$ particles using systematic resampling with uniform draw $\bar{u}_{t+1}$ and weights $w_{t_{i+1}}^{(k)}$, $k=1,\ldots,N$.
\end{enumerate}

Output: $N$ particles $\{n_{t+1}^{(k)}, k=1,\ldots,N\}$ to be used in step $t+1$, an estimate for the current marginal likelihood term $\hat{\pi}_{u_{t+1}}(y_{t+1}|y_{1:t},\theta)=\frac{1}{N}\sum_{k=1}^{N}\tilde{w}_{t+1}^{(k)}$.
\end{algorithm}

Pseudo-marginal Metropolis-Hastings methods \citep[PMMH,][]{andrieu09,andrieu09b} are a class of Metropolis-Hastings (MH) scheme that target the joint density
\[
\pi(u,\theta) \propto \pi(\theta) g(u) \hat{\pi}_{u}(y|\theta)
\]
for which it is easily checked that marginalising over $U$ gives the marginal parameter posterior $\pi(\theta|y)$. Hence, an MH scheme with proposal density $q(\theta^*|\theta)g(u^*)$ and acceptance probability 
\[
\alpha\left(\{\theta^*,u^*\}|\{\theta,u\}\right) = \textrm{min} \left\{ 1, \frac{\pi(\theta^*) \hat{\pi}_{u^*}(y|\theta^*)}{\pi(\theta) \hat{\pi}_{u}(y|\theta)} \times \frac{q(\theta|\theta^*)}{q(\theta^*|\theta)} \right\}
\]
targets the joint density $\pi(u,\theta)$ for with retaining draws of $\theta$ gives (dependent) samples from the marginal parameter posterior. 

The efficiency of the PMMH scheme can be improved by proposing the auxiliary variable from a $g$-reversible kernel $f(u^*|u)$ that induces positive correlation between $u$ and $u^*$, and in turn, $\hat{\pi}_{u}(y|\theta)$ and $\hat{\pi}_{u^*}(y|\theta^*)$, so that the variance of the acceptance probability is reduced. Suppose that $g(u)=\textrm{N}\left(u;\,0\,,\,I_{\textrm{dim}(u)}\right)$ and note that where necessary, the inverse CDF method can be used to transform the auxiliary variable to be Gaussian (for example, in step 3 of Algorithm~\ref{auxPF}) where a uniform draw is required). A practical choice of $f(u^*|u)$ is the $g$-reversible Crank-Nicolson kernel
\[
f(u^*|u)=\textrm{N}\left(u^*;\,\rho u\,,\,\left(1-\rho^2\right)I_{\textrm{dim}(u)}\right)
\]
for which the tuning parameter $\rho$ controls correlation between $u$ and $u^*$. The resulting correlated PMMH scheme \citep[CPMMH,][]{dahlin2015,deligiannidis2018} is an MH scheme targeting $\pi(u,\theta)$ with proposal density $q(\theta^*|\theta)f(u^*|u)$ and acceptance probability as above. CPMMH can result in significant gains in computational efficiency over PMMH \citep[see e.g.][in the context of stochastic kinetic models]{GoliBrad19}, provided that the positive correlation between $u$ and $u^*$ induces positive correlation between successive likelihood estimates. To alleviate the issue of resampling in the particle filter potentially eroding this correlation, we follow \cite{choppala2016} by sorting particles (according to Euclidean distance from the particle with the smallest first component) before propagation.     

Finally we note that when interest lies in the posterior for the latent incidence process, samples can be obtained via modification of the (C)PMMH scheme \citep{andrieu10} by drawing a particle path at each iteration of the algorithm. Note that this requires storing the ancestral lineages of the particles in each run of the particle filter. 

\subsubsection{Analytic method via Gaussian approximation}
\label{sec:ffbs}

The LNA, when combined with the linear Gaussian observation model (\ref{obsNorm}), permits analytic calculation of the observed data likelihood $\pi(y|\theta)$ and the conditional posterior $\pi(n|\theta,y)$. Evaluation of the former can be efficiently achieved via a forward filter and draws from the latter via backward sampling. We apply these methods to the Binomial and Negative Binomial observation models in (\ref{obsBin}) and (\ref{obsNBin}) through suitable Gaussian approximations thereof. For reasons of brevity, we focus on the Binomial case but note that our approach is easily extended to the Negative Binomial case. Where appropriate, we suppress the parameter vector $\theta$ from the notation for simplicity.

To make clear the two appoximations to be used in the filtering recursions, consider the LNA written in state-space format over a time interval $(t,t+1]$, using a Gaussian approximation to the Binomial observation model. We have that 
\begin{align}
N_{t+1}|(N_t=n_t) &\sim \textrm{N}\left(\eta_{t+1}+G_{t+1}(n_t-\eta_t), V_{t+1} \right), \label{lnaSS1}\\
Y_{t+1}|(N_{t+1}=n_{t+1},N_t=n_t) &\sim \textrm{N}\left(\lambda P'\Delta n_{t+1}, 
\lambda(1-\lambda)P'\Delta {n}_{t+1}\right), \label{lnaSS2}
\end{align}  
where $\eta_{t+1}$, $G_{t+1}$ and $V_{t+1}$ are obtained by integrating (\ref{lna1}), (\ref{lna2}) and (\ref{lna3}) over $(t,t+1]$ 
with initial conditions of $\eta_t$ (itself integrated from time $0$), $I_2$ and $0_2$. Although (\ref{lnaSS1}) is linear 
in $n_t$, as noted in Section~\ref{sec:pmmh}, $\eta_t$ should be initialised at $n_t$. `Restarting' 
the LNA in this way can avoid issues arising from the ODE solution becoming poor over long time intervals \citep{fearnhead14,minas17}. However, we now have that both (\ref{lnaSS1}) and (\ref{lnaSS2}) involve nonlinear expressions of the latent process. Therefore, to permit the use of standard Kalman-filtering recursions we make further linear approximations. Suppose that the filtering distribution at time $t$ is $N_t|(Y_{1:t}=y_{1:t})\sim\textrm{N}(a_t,C_t)$. Firstly, we set $\eta_t=a_t$, $V_t=C_t$ and integrate (\ref{lna1}) and (\ref{lna3}) over $(t,t+1]$ to obtain $\eta_{t+1}$ and $V_{t+1}$. Finally, we replace $\Delta n_{t+1}$ in the variance of (\ref{lnaSS2}) with $\Delta \widehat{n}_{t+1}:=\textrm{E}(\Delta N_{t+1})=\eta_{t+1}-a_t$. Note that explicit conditioning of the expectation on $y_{1:t}$ has been supressed for simplicity. The resulting linear and Gaussian state-space model is
\begin{align}
N_{t+1}|(N_t=n_t) &\sim \textrm{N}\left(\eta_{t+1}, V_{t+1} \right), \label{lnaSS3}\\
Y_{t+1}|(N_{t+1}=n_{t+1},N_t=n_t) &\sim \textrm{N}\left(\lambda P'\Delta n_{t+1}, 
\lambda(1-\lambda)P'\Delta \widehat{n}_{t+1}\right). \label{lnaSS4}
\end{align}         
In the remainder of this section, we derive the filtering recursions based on (\ref{lnaSS3}) and (\ref{lnaSS4}) to compute the observed data likelihood and conditional posterior of the latent process.
 
We construct the observed data likelihood contribution $\pi(y_{t+1}|y_{1:t},\theta)$ as follows. Conditional on $y_{1:t}$, we have that 
\[
\textrm{Var}(\Delta N_{t+1}) = V_{t+1} + C_t - C_t G_{t+1}' - G_{t+1}C_t
\]  
where we have used that $\textrm{Cov}(N_{t+1},N_{t})=G_{t+1}\textrm{Var}(N_t)$. Hence, combining with (\ref{lnaSS4}) gives
\begin{align}
\pi(y_{t+1}|y_{1:t},\theta)&=
\textrm{N}\left(y_{t+1}; \lambda P'\textrm{E}(\Delta N_{t+1})\,,\, \lambda^2 P' \textrm{Var}(\Delta N_{t+1})P +\widehat{\sigma}^2 \right) \label{margllLNA}
\end{align}
where $\widehat{\sigma}^2=\lambda(1-\lambda) P'\Delta\widehat{n}_{t+1}$ is the observation variance in (\ref{lnaSS4}). To update the filtering distribution, we construct the joint density of $N_{t+1}$ and $Y_{t+1}$ conditional on $Y_{1:t}=y_{1:t}$ as
\[
\begin{pmatrix}
N_{t+1} \\ 
Y_{t+1} 
\end{pmatrix} \sim \textrm{N}\left\{   
\begin{pmatrix}
\eta_{t+1} \\
\lambda P'\textrm{E}(\Delta N_{t+1})
\end{pmatrix}\,,\,
\begin{pmatrix}
V_{t+1} & \textrm{Cov}(N_{t+1},Y_{t+1})\\
\textrm{Cov}(Y_{t+1},N_{t+1}) & \lambda^2 P' \textrm{Var}(\Delta N_{t+1})P +\widehat{\sigma}^2
\end{pmatrix}
\right\}
\]
where $\textrm{Cov}(N_{t+1},Y_{t+1})=\lambda (V_{t+1} - G_{t+1}C_t)P$. Hence, conditioning on $Y_{t+1}=y_{t+1}$ gives\\ $N_{t+1}|\left(Y_{1:t+1}=y_{1:t+1}\right)\sim\textrm{N}(a_{t+1},C_{t+1})$ with mean
\begin{align}
a_{t+1} &= \eta_{t+1} + \textrm{Cov}(N_{t+1},Y_{t+1})(\lambda^2 P' \textrm{Var}(\Delta N_{t+1})P +\widehat{\sigma}^2)^{-1}(y_{t+1}-\lambda P'\textrm{E}(\Delta N_{t+1}) ) \label{LNAa}
\end{align}
and variance
\begin{align}
C_{t+1} &= V_{t+1} - \textrm{Cov}(N_{t+1},Y_{t+1})(\lambda^2 P' \textrm{Var}(\Delta N_{t+1})P +\widehat{\sigma}^2)^{-1}\textrm{Cov}(Y_{t+1},N_{t+1}). \label{LNAC}
\end{align}
Calculation of (\ref{margllLNA}), (\ref{LNAa}) and (\ref{LNAC}) constitutes a single step of the forward filter; see Algorithm~\ref{algLNAff}, which can be iterated over $t$ to give an evaluation of the observed data likelihood (under the LNA), $\pi(y|\theta)$. Hence, draws from the marginal parameter posterior $\pi(\theta|y)$, with the LNA as the inferential model, are obtained in a straightforward manner via Metropolis-Hastings e.g. random walk Metropolis (RWM). 

\begin{algorithm}[t]
\caption{Step $t+1$ of the LNA Forward Filter}\label{algLNAff}
Input: Parameter $\theta$; $a_{t}$ and $C_{t}$, the initial conditions of (\ref{lna1}) and (\ref{lna3}); $\pi(y_{1:t}|\theta)$, the current observed data likelihood; $y_{t+1}$, the next observation.
\begin{enumerate}
\item Prior at $t+1$. Initialise the LNA with $\eta_{t}=a_{t}$, $G_t=1_2$ and $V_{t}=C_{t}$. Integrate (\ref{lna1}), (\ref{lna2}) and (\ref{lna3}) forward to $t+1$ to obtain $\eta_{t+1}$, $G_{t+1}$ and $V_{t+1}$. Thus
\[
N_{t+1}|\left(Y_{1:t}=y_{1:t}\right) \sim \textrm{N}(\eta_{t+1},V_{t+1}).
\]
\item Likelihood update. Compute
\[
\pi(y_{1:t+1}|\theta) = \pi(y_{1:t}|\theta)\pi(y_{t+1}|y_{1:t},\theta)
\]
where $\pi(y_{t+1}|y_{1:t},\theta)$ is given by (\ref{margllLNA}).
\item Posterior at $t+1$. Combining the distributions of $N_{t+1}$ and $Y_{t+1}$ (given $y_{1:t}$) and then conditioning on $y_{t+1}$ gives $N_{t+1}|\left(Y_{1:t+1}=y_{1:t+1}\right) \sim \textrm{N}(a_{t+1},C_{t+1})$ where $a_{t+1}$ and $C_{t+1}$ are given by (\ref{LNAa}) and (\ref{LNAC}).
\end{enumerate}
Output: $\pi(y_{1:t+1}|\theta)$, $a_{t+1}$ and $C_{t+1}$.
\end{algorithm}

It remains that we can use the LNA to generate draws from $\pi(n|\theta,y)$. Note the factorisation
\[
\pi(n|\theta,y)=\prod_{t=1}^{T-1} \pi(n_t|n_{t+1},y_{1:t},\theta)
\]
where each constituent term is a Gaussian density. Under the LNA, the joint density of $N_t$ and $N_{t+1}$ conditional on $Y_{1:t}=y_{1:t}$ is 
\[
\begin{pmatrix}
	N_{t} \\	
	N_{t+1}
	\end{pmatrix}\sim \textrm{N}\left\{\begin{pmatrix}
	a_{t} \\[0.2em]	
	\eta_{t+1} 	
	\end{pmatrix}\,,\, \begin{pmatrix}
	C_{t} & C_{t}G_{t+1}'   \\[0.2em]	
	G_{t+1}C_{t} & V_{t+1}  	 
	\end{pmatrix} \right \}. 
\]  
Conditioning on $N_{t+1}=n_{t+1}$ gives $N_t|\left(N_{t+1}=n_{t+1},Y_{1:t}=y_{1:t},\theta\right)\sim \textrm{N}(\tilde{a}_t,\tilde{C}_t)$ with mean and variance 
\begin{align*}
\tilde{a}_{t} &= a_{t}+C_{t}G_{t+1}' V_{t+1}^{-1}\left(n_{t+1}-\eta_{t+1}\right), \\
\tilde{C}_{t} &= C_{t}-C_{t}G_{t+1}' V_{t+1}^{-1}G_{t+1}C_{t}.
\end{align*}
Hence, the components of the cumulative incidence $n$ can be drawn via backward sampling for $t=T,T-1,\ldots,1$, given storage of the LNA ODE output and filtering mean/variance from the forward filter. Then, the latent process $x$ can be constructed deterministically from $n$ and the initial values $x_0$ using (\ref{MJP1}). 

\section{Applications}
\label{sec:app}

We consider two applications of the methodology described in Section~\ref{sec:binf}. Firstly, using synthetic data generated from the SIR model, we compare the performance of the two marginalisation techniques described in Sections~\ref{sec:pmmh} and \ref{sec:ffbs}; these are the correlated pseudo-marginal Metropolis-Hastings (CPMMH) and forward filtering Metropolis-Hastings (henceforth FFMH) based inference schemes. We  compare the accuracy of posterior output from these schemes with inferences obtained by assuming the most natural Markov jump process as the inferential model. We fit this model using the pseudo-marginal Metroplis-Hastings (PMMH) scheme described in \cite{GoliWilk11}. Additionally, we include inferences based on a deterministic ODE model of latent incidence (fit via MH). In the second application, we use FFMH to fit SIR and SIRS models wth two different choices of observation model, to a real data set consisting of pest removals from trees in a London park.    

All algorithms are coded in R and were run on a desktop computer with an Intel quad-core CPU. Source code is available at \url{https://github.com/AndyGolightly/LNA-Incidence}. All schemes use random walk proposals with Gaussian innovations for the log-transformed parameters. For CPMMH, we fixed $\rho=0.99$, which we found to give a good balance between mixing over the auxiliary variable and parameter chains. We chose the number of particles $N$ by following the practical advice of \cite{deligiannidis2018}. That is, we choose $N$ so that the variance of $\log \hat{\pi}_{u^*}(y|\theta^*) - \log \hat{\pi}_{u}(y|\theta)  \approx 1$. For CPMMH and FFMH, we took the random walk innovation variance to be $\widehat{\textrm{Var}}(\log\theta|y)$ estimated from a pilot run, and subsequently scaled to meet a desired empirical acceptance rate (see e.g. \cite{schmon21} for (C)PMMH and \cite{schmon22} for Metropolis-Hastings). CPMMH was run for $50,000$ iterations and the remaining schemes were run for $10,000$ iterations, which we found gave reasonable mixing efficiency as measured by, for example, effective sample size \citep[see e.g.][]{Plummer06}. 

\subsection{Simulation study}
\label{sec:ss}

\begin{figure}[ht!]
\centering
\psfragscanon
\psfrag{time}[][][0.7][0]{$t$ (days)}
\psfrag{Yt}[][][0.7][-90]{$Y_t$}
\psfrag{St}[][][0.7][-90]{$S_t$}
\psfrag{It}[][][0.7][-90]{$I_t$}
\includegraphics[width=5.0cm,height=16cm,angle=-90]{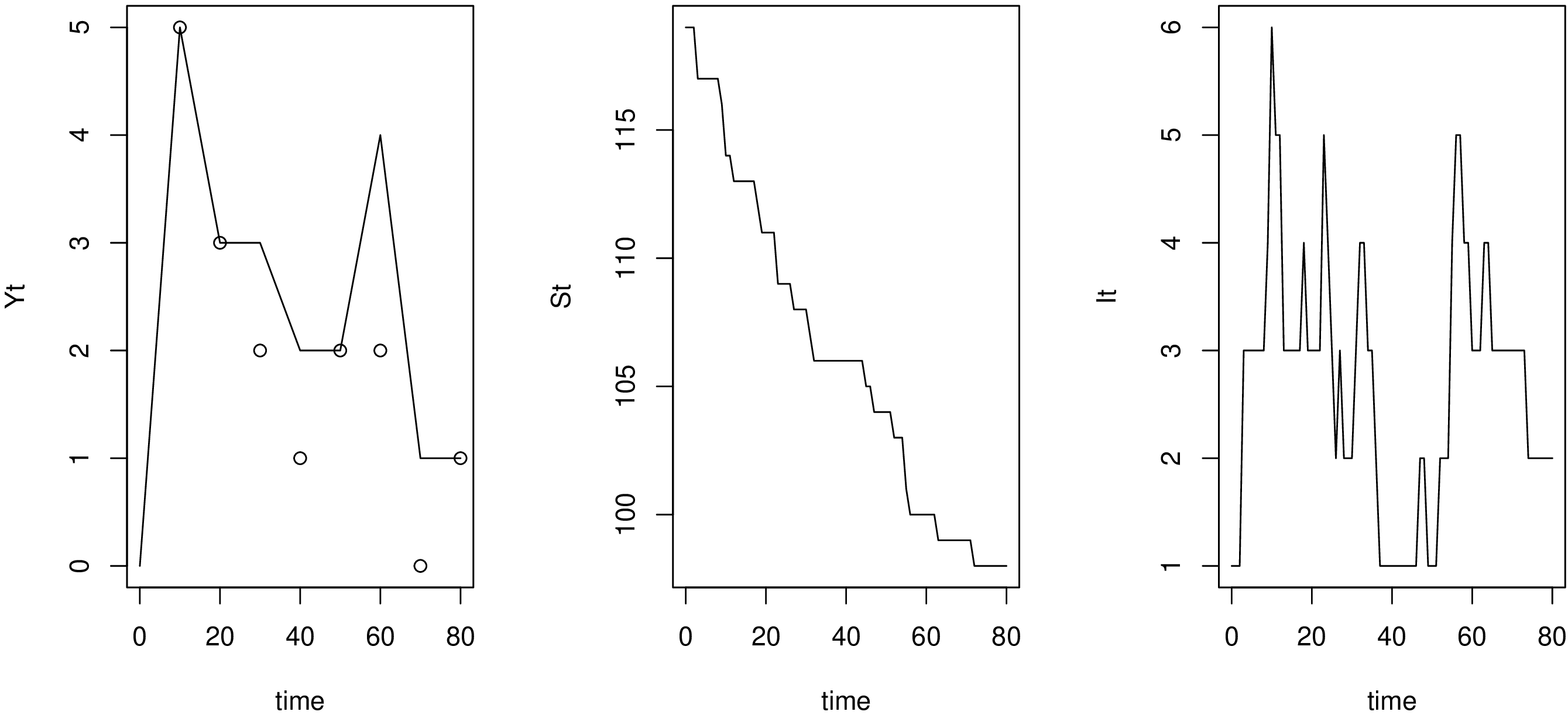}
\includegraphics[width=5.0cm,height=16cm,angle=-90]{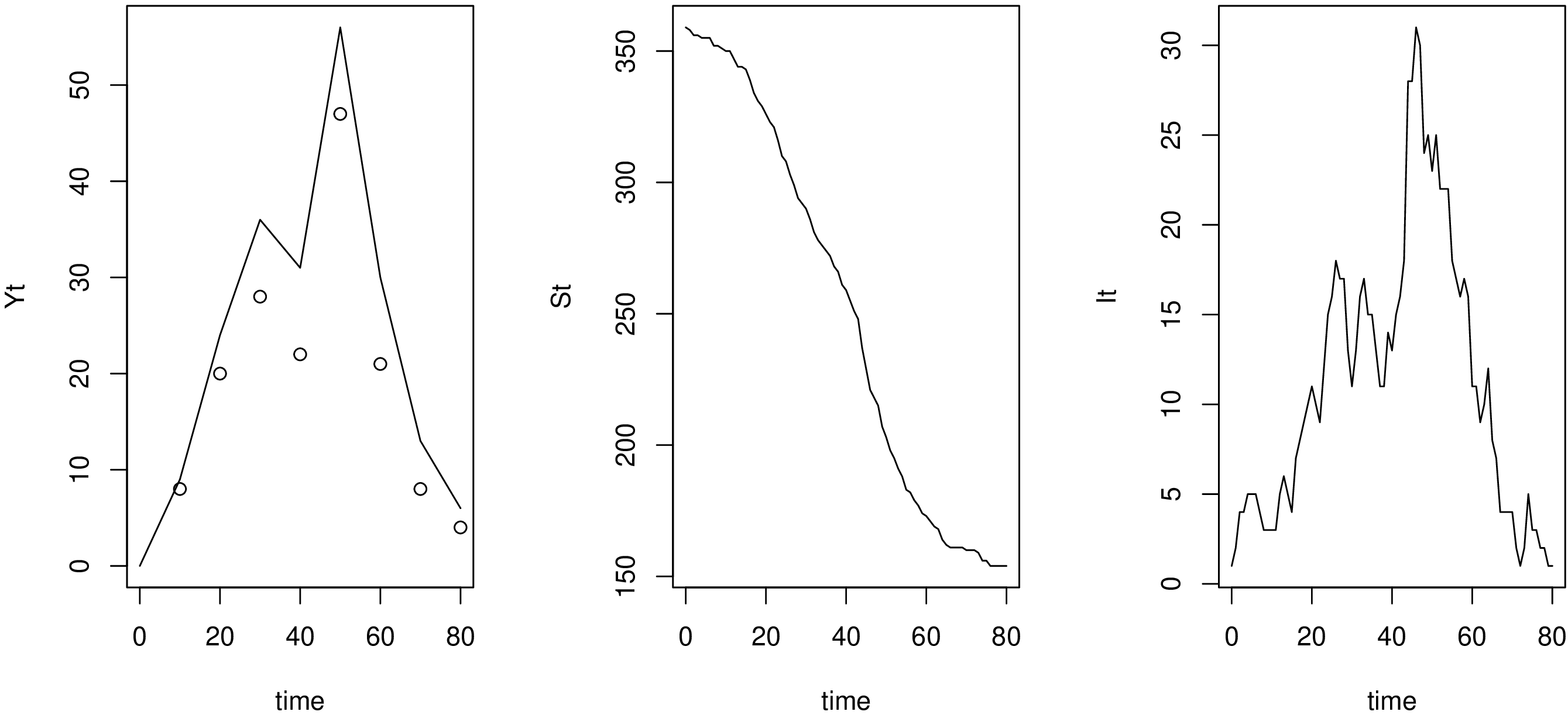}
\includegraphics[width=5.0cm,height=16cm,angle=-90]{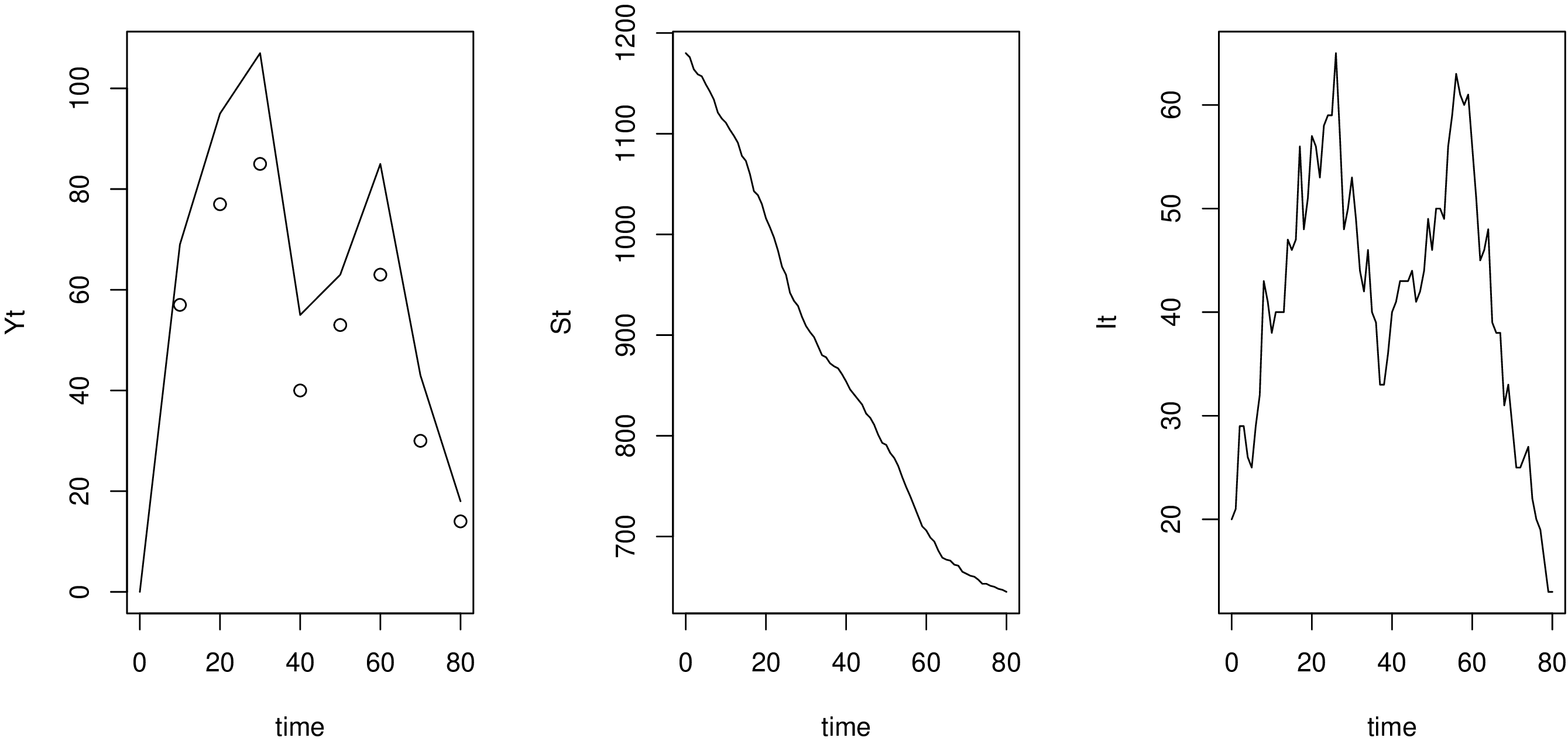}
\caption{Synthetic data sets $\mathcal{D}_1$ (top panel), $\mathcal{D}_2$ (middle panel) and $\mathcal{D}_3$ (bottom panel). Left: noisy numbers of new infecteds in a 10 day interval (circles) and latent values (line). Middle and right: corresponding susceptible and infected states.}
\label{fig:SSdata}
\end{figure}

We generated three synthetic data sets (denoted $\mathcal{D}_i$, $i=1, 2, 3$) from the SIR model, each consisting of the number of new infections in time intervals $(t,t+10]$ for $t=10,20,\ldots,70$. For $\mathcal{D}_1$, we used $x_0=(119,1)'$ and $(\beta,\gamma)'=(0.00091,0.082)'$; these choices are consistent with inferences from the well studied Abakaliki small pox data \citep[see e.g.][]{bailey1975}. For $\mathcal{D}_2$, we constructed a larger outbreak by scaling the total population size $N_{\textrm{pop}}$ and removal rate by a factor of 3, resulting in $x_0=(359,1)'$ and $(\beta,\gamma)'=(0.00091,0.246)'$. For $\mathcal{D}_3$, we scaled $N_{\textrm{pop}}$ by a factor of 10 (compared to $\mathcal{D}_1$) and set $x_0=(1180,20)'$. We scaled both the infection and removal rates to give $(\beta,\gamma)'=(0.00018,0.164)'$. Note that all data sets have the same basic reproduction number, $\mathcal{R}_0=N_{\textrm{pop}}\beta/\gamma =1.33$. We corrupted the resulting incidences via the Binomial observation model (\ref{obsBin}) with $\lambda=0.8$ in each case. The data sets are shown in Figure~\ref{fig:SSdata} alongside the underlying traces of $S_t$ and $I_t$ (assumed unobserved).  

When analysing $\mathcal{D}_1$, we adopted an independent prior specification with Gamma and Uniform components by taking $\beta\sim \textrm{Gamma}(10,10^4)$, $\gamma\sim \textrm{Gamma}(10,10^2)$ and $\lambda\sim \textrm{Unif}(0,1)$. The first two choices have been used by \cite{fearnhead2004} and many others when analysing the Abakaliki small pox data. For $\mathcal{D}_2$ and $\mathcal{D}_3$ we adopted a more diffuse prior for the removal rate to better reflect the increase in the ground truth value; specifically, $\gamma\sim \textrm{Gamma}(10,30)$, with the prior specification for the remaining components as for $\mathcal{D}_1$. We assume that the initial state $x_0$ is fixed and known but note that inference for $x_0$ is possible by augmenting $\theta$ to include the components of $x_0$.   

\begin{table}[ht!]
\centering
\small
	\begin{tabular}{@{}l lll llll@{}}
         \toprule
Model / Scheme  & $\rho$ & $N$ & mESS/s & \multicolumn{4}{l}{Mean (Std. Dev.)}  \\
\cmidrule(l){5-8}
        &         &     &      & $\beta$ & $\gamma$ & $\lambda$ & $\mathcal{R}_0$\\   
\midrule
 & & & & \multicolumn{4}{c}{Data set $\mathcal{D}_1$}  \\
 & & & & 0.00091 & 0.082 & 0.8 & 1.33 \\
MJP / PMMH  &  0.00 & \phantom{0}30                     & \phantom{0}0.682 & 0.00091 (0.00024)& 0.088 (0.022) & 0.62 (0.22) & 1.29 (0.35)\\
LNA / PMMH  &  0.00 & \phantom{0}25                     & \phantom{0}0.039 & 0.00107 (0.00027)& 0.094 (0.025) & 0.61 (0.24) & 1.37 (0.41)\\
LNA / CPMMH &  0.99 & \phantom{0}15                     & \phantom{0}0.064 & 0.00111 (0.00024)& 0.101 (0.027) & 0.64 (0.22) & 1.36 (0.40)\\
LNA / FFMH  &  \phantom{0}-- & \phantom{0}--            & \phantom{0}3.075 & 0.00102 (0.00021)& 0.117 (0.029) & 0.82 (0.15) & 1.09 (0.26)\\
ODE / MH    &  \phantom{0}-- & \phantom{0}--            & 12.091           & 0.00203 (0.00024)& 0.192 (0.042) & 0.48 (0.16) & 1.35 (0.42)\\
& & & & \multicolumn{4}{c}{Data set $\mathcal{D}_2$}  \\
 & & & & 0.00091 & 0.246 & 0.8 & 1.33 \\
MJP / PMMH  &  0.00 & 125                               & \phantom{0}0.091  & 0.00087 (0.00016)& 0.225 (0.041)& 0.75 (0.12) & 1.35 (0.19)\\
LNA / PMMH  &  0.00 & 120                               & \phantom{0}0.018  & 0.00092 (0.00017)& 0.231 (0.050)& 0.77 (0.12) & 1.51 (0.23)\\
LNA / CPMMH &  0.99 & \phantom{0}60                     & \phantom{0}0.030  & 0.00089 (0.00016)& 0.228 (0.056)& 0.77 (0.12) & 1.44 (0.25)\\
LNA / FFMH  &  \phantom{0}-- & \phantom{0}--            & \phantom{0}3.680  & 0.00094 (0.00021)& 0.234 (0.060)& 0.78 (0.12) & 1.49 (0.22)\\
ODE / MH    &  \phantom{0}-- & \phantom{0}--            & 13.440            & 0.00087 (0.00005)& 0.188 (0.021)& 0.65 (0.04) & 1.68 (0.10)\\ 
 & & & & \multicolumn{4}{c}{Data set $\mathcal{D}_3$}  \\
 & & & & 0.00018 & 0.164 & 0.8 & 1.33 \\
MJP / PMMH  &  0.00 & 120                               & \phantom{0}0.045  & 0.00032 (0.00006)& 0.290 (0.059)& 0.82 (0.09) & 1.33 (0.10)\\
LNA / PMMH  &  0.00 & 120                               & \phantom{0}0.012  & 0.00034 (0.00007)& 0.318 (0.070)& 0.83 (0.10) & 1.30 (0.11)\\
LNA / CPMMH &  0.99 & \phantom{0}60                     & \phantom{0}0.019  & 0.00034 (0.00006)& 0.308 (0.062)& 0.81 (0.12) & 1.32 (0.12)\\
LNA / FFMH  &  \phantom{0}-- & \phantom{0}--            & \phantom{0}3.533  & 0.00023 (0.00005)& 0.217 (0.049)& 0.85 (0.10) & 1.29 (0.11)\\
ODE / MH    &  \phantom{0}-- & \phantom{0}--            & 12.984            & 0.00019 (0.00001)& 0.159 (0.008)& 0.63 (0.01) & 1.45 (0.04)\\ 

\bottomrule
\end{tabular}
      \caption{Synthetic data application. Inferential model / scheme, correlation parameter, number of particles, minimum effective sample size per second and marginal parameter posterior summaries. The ground truth parameter values are indicated for each data set.}\label{tab:tabEpi1}	
\end{table}

\begin{figure}[ht!]
\centering
\psfragscanon
\psfrag{Be}[][][0.7][0]{$\beta$}
\psfrag{Ga}[][][0.7][0]{$\gamma$}
\psfrag{logitL}[][][0.7][0]{logit $\lambda$}
\psfrag{R0}[][][0.7][0]{$\mathcal{R}_0$}
\includegraphics[width=6.0cm,height=17cm,angle=-90]{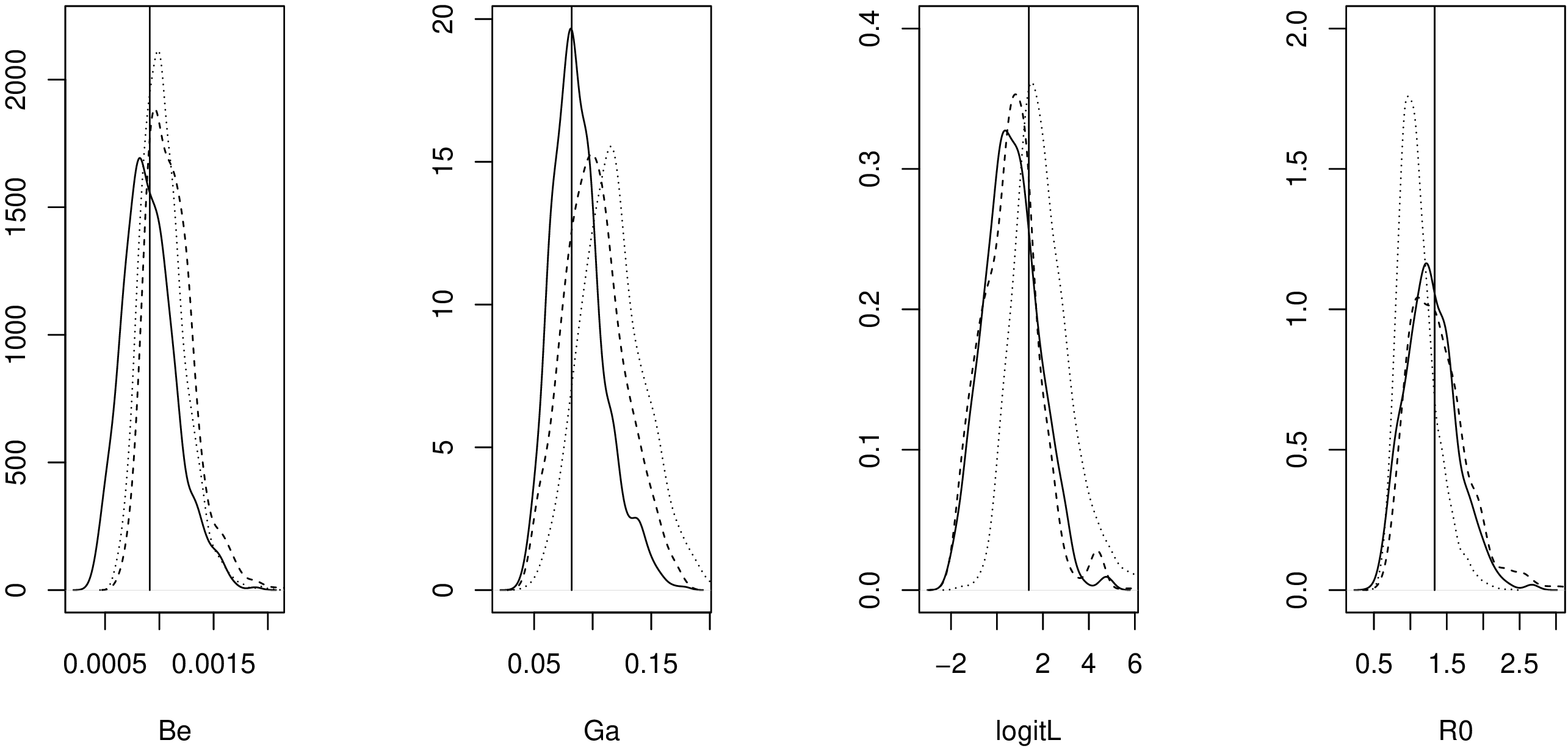}
\includegraphics[width=6.0cm,height=17cm,angle=-90]{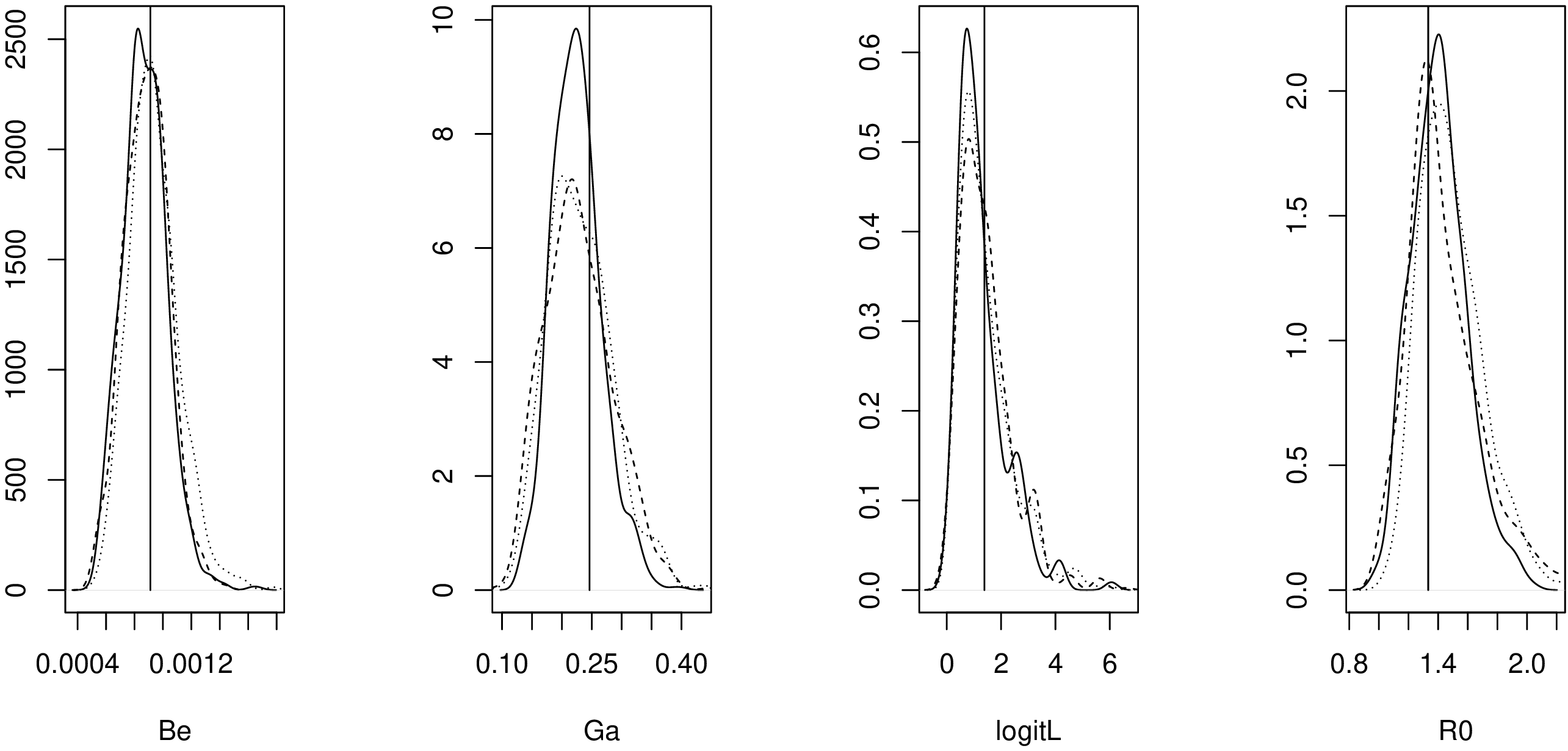}
\includegraphics[width=6.0cm,height=17cm,angle=-90]{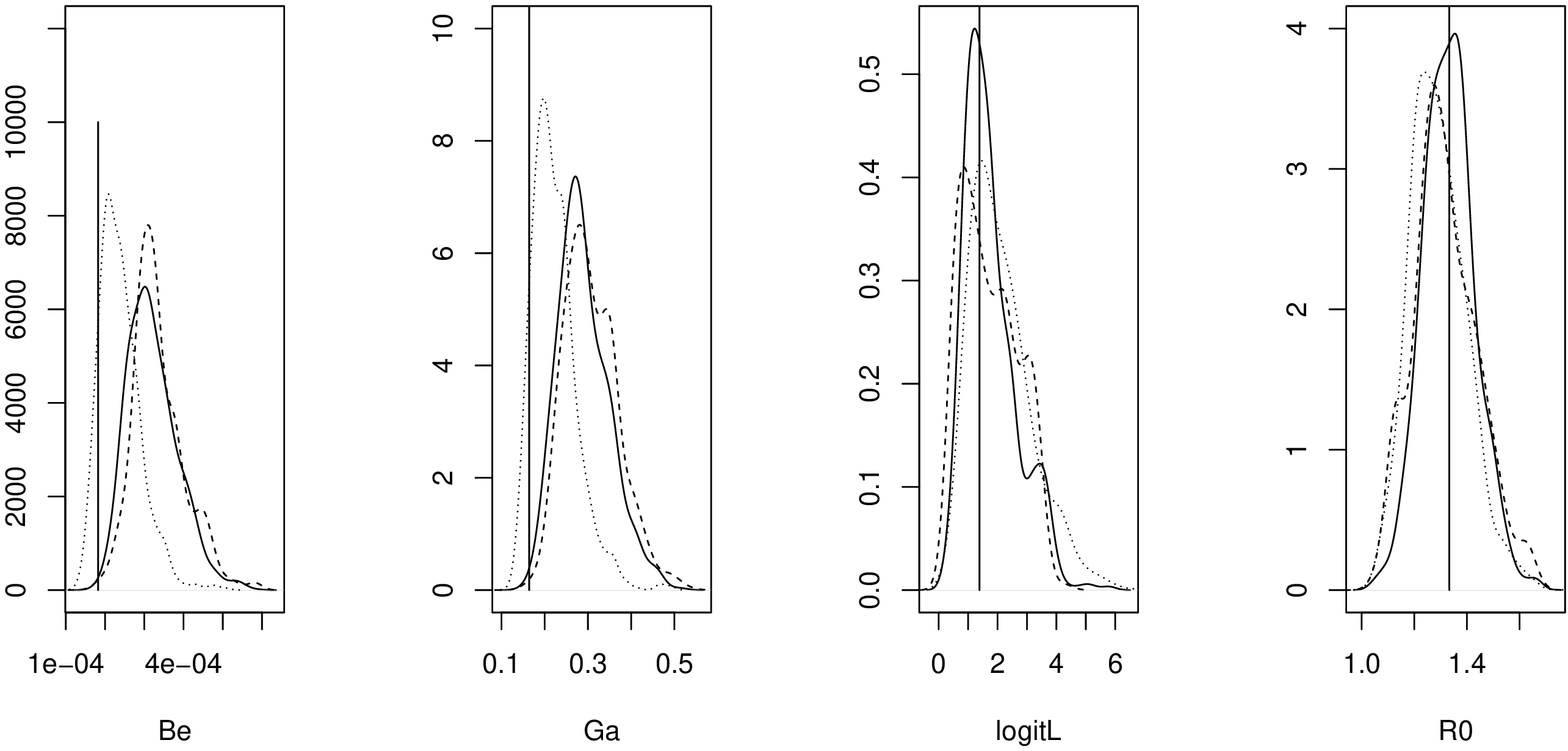}
\caption{Synthetic data application. Marginal posterior densities based on $\mathcal{D}_1$ (top panel), $\mathcal{D}_2$ (middle panel) and $\mathcal{D}_3$ (bottom panel), and using the output of MJP / PMMH (solid line), LNA / CPMMH (dashed line), LNA / FFMH (dotted line).}
\label{fig:SSpost}
\end{figure}

\begin{figure}[ht!]
\centering
\psfragscanon
\psfrag{time}[][][0.7][0]{$t$ (days)}
\psfrag{St}[][][0.7][-90]{$S_t$}
\psfrag{It}[][][0.7][-90]{$I_t$}
\includegraphics[width=5.5cm,height=16cm,angle=-90]{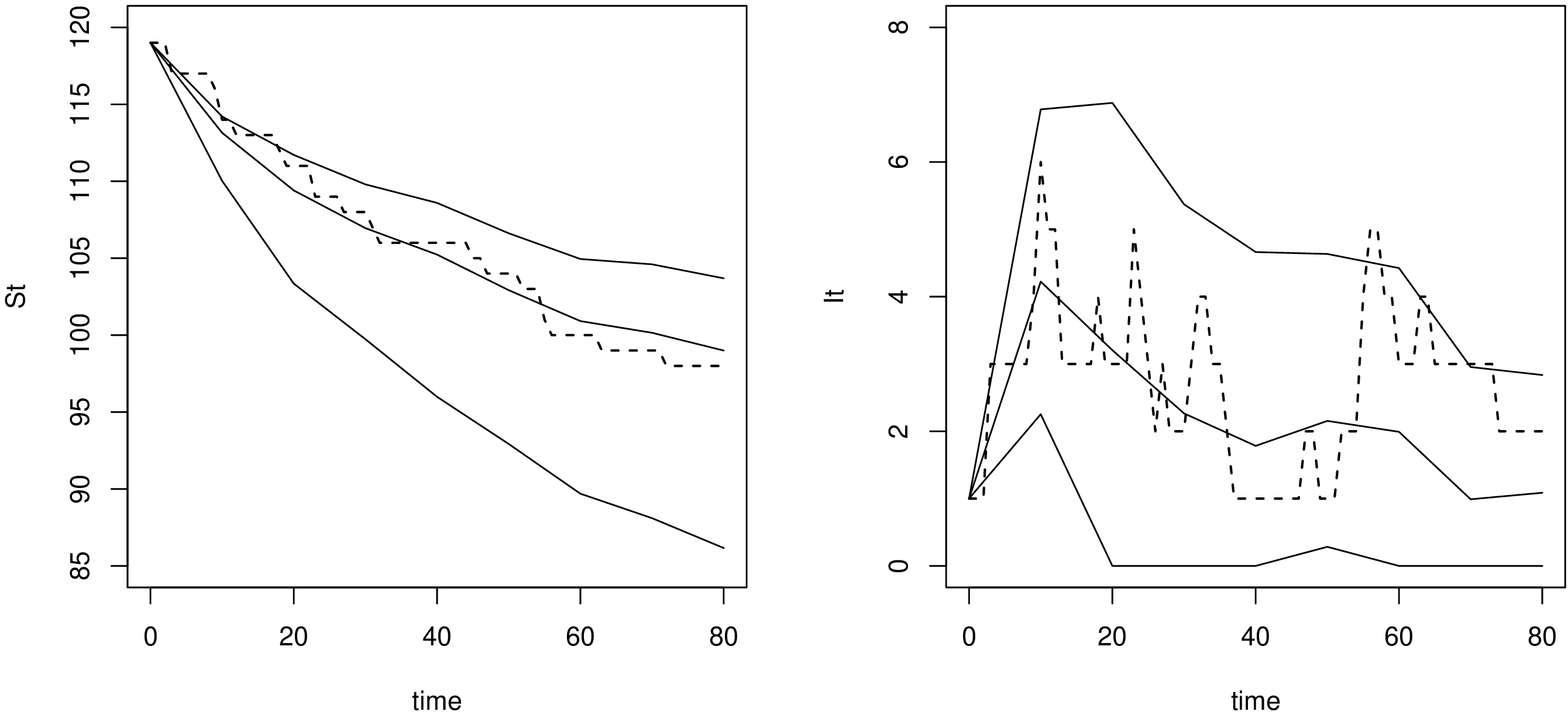}
\includegraphics[width=5.5cm,height=16cm,angle=-90]{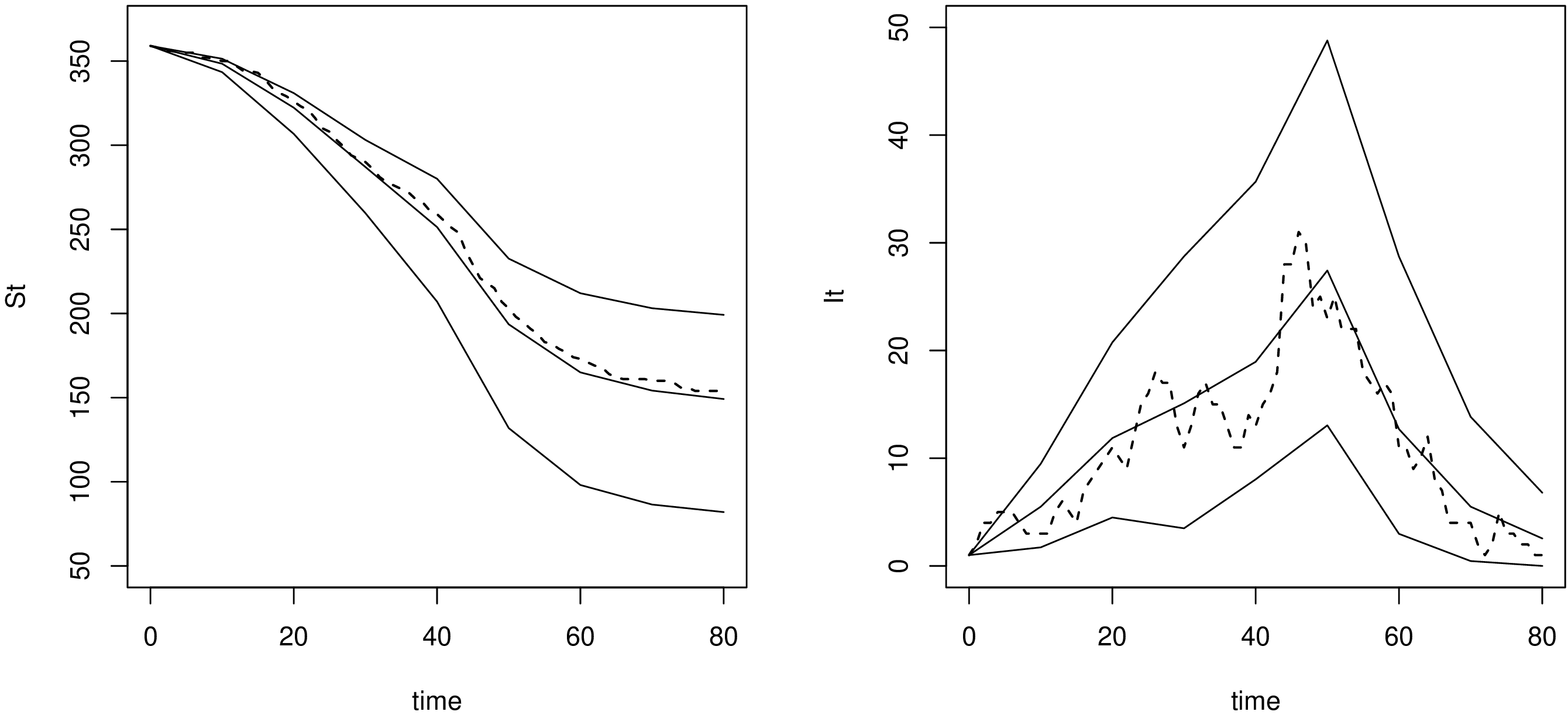}
\includegraphics[width=5.5cm,height=16cm,angle=-90]{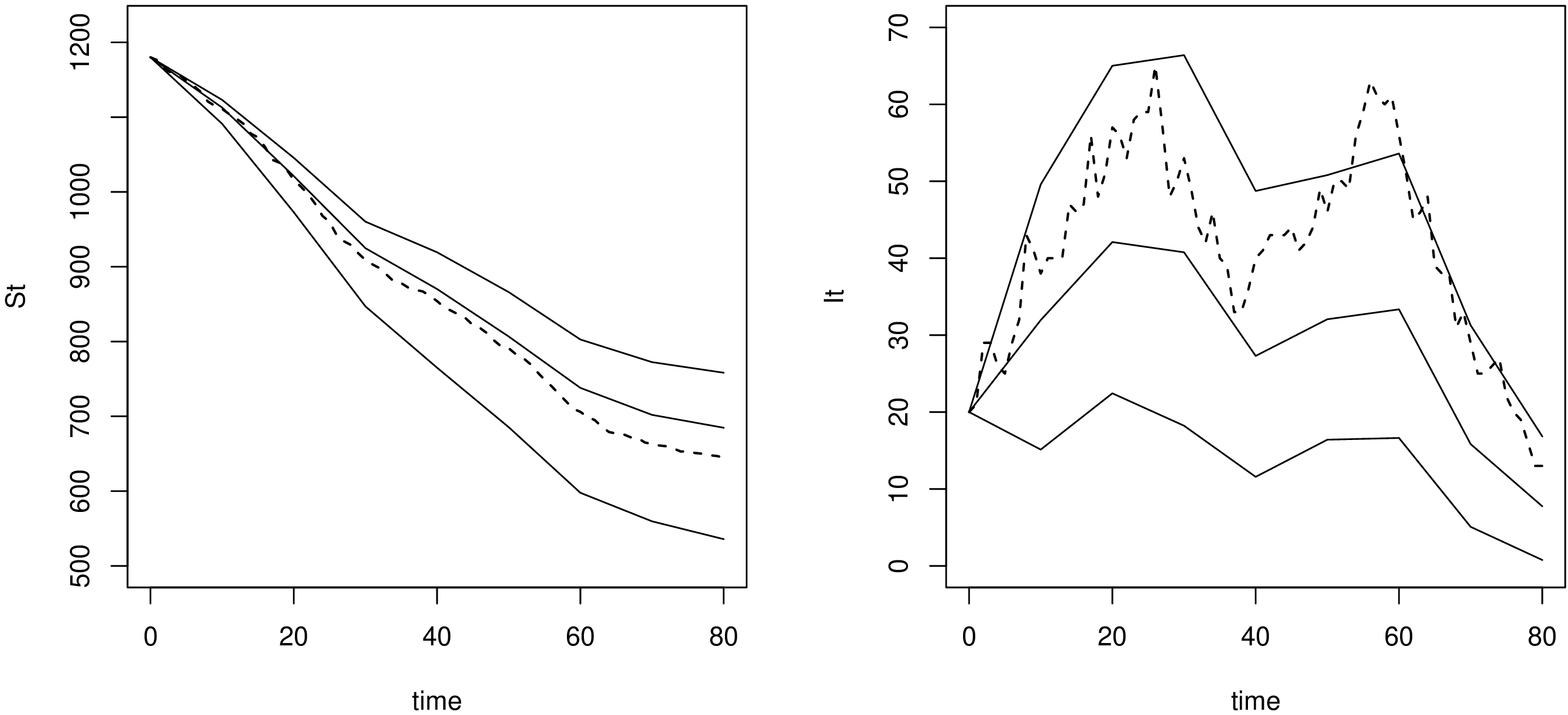}
\caption{LNA within-sample predictive distributions (mean and 95\% credible intervals) for $S_t$ (left) and $I_t$ (right) based 
on synthetic data sets $\mathcal{D}_1$ (top panel), $\mathcal{D}_2$ (middle panel) and $\mathcal{D}_3$ (bottom panel). Ground truth latent process trajectories are shown as dashed lines.}
\label{fig:SSpred}
\end{figure}

Table~\ref{tab:tabEpi1} and Figures~\ref{fig:SSpost}--\ref{fig:SSpred} summarise the posterior output from each scheme. For data set $\mathcal{D}_1$ (population size $N_{\textrm{pop}}=120$), although all inferential models give posterior output that is consistent with the ground truth parameter values, there are noticeable inconsistencies between LNA- and MJP-based inferences. Using the LNA to model the latent incidence process but with the correct observation model (LNA / (C)PMMH) results in an overestimation of the infection and removal rates, although there is little difference between this approach compared to the MJP when considering the basic reproduction number $\mathcal{R}_0$. However, differences are more pronounced when further approximating the observation model as Gaussian (LNA / FFMH), which results in under estimation of $\mathcal{R}_0$. Nevertheless, these differences are relatively small, and the advantages (in terms of overall efficiency) of analytically integrating out the latent process (as per LNA / FFMH) are clear. 

We measure overall efficiency using minimum (over each parameter chain) effective sample size (ESS) per second (mESS/s). Given the small population size for $\mathcal{D}_1$, using the MJP inside a PMMH scheme is computationally more efficient than using the LNA (which, as implemented, requires numerical integration of 5 coupled ODEs per particle per iteration). Correlating successive likelihood estimates (LNA / CPMMH vs PMMH) increases overall efficiency by a factor of 2, however, the largest gains in overall efficiency are obtained by LNA / FFMH, which improves on MJP / PMMH by a factor of 4 and on LNA / CPMMH by a factor of almost 50.

For data set $\mathcal{D}_2$ (population size $N_{\textrm{pop}}=360$), the pseudo-marginal schemes require more particles, due to the intrinsic stochasticity of realisations of the latent process generated inside the particle filter, which is large compared to observation noise. The increased population size (and corresponding parameter values that generated the data) leads to many more reaction occurrences between observation instants (compared to $\mathcal{D}_1$) reducing the relative efficiency of MJP / PMMH versus LNA / (C)PMMH and LNA / FFMH. We note that for this data set, using the LNA to model the latent process and additionally taking a linear Gaussian approximation of the observation model leads to an inference scheme that is both efficient (with an mESS/s 40 times larger than that of the next best perfoming scheme) and accurate (see Figure~\ref{fig:SSpost}, bottom panel). 

The magnitude of of typical observations in data set $\mathcal{D}_3$ (population size $N_{\textrm{pop}}=1200$) is broadly consistent with that of $\mathcal{D}_2$ and we find that the pseudo-marginal schemes require similar particle numbers. Using the LNA with the correct observation model gives parameter inferences that are consistent with the MJP. Using LNA / FFMH appears to result in underestimates of the infection and removal rates although the basic reproduction number appears to be accurately estimated. In terms of overall efficiency, the advantage of LNA / FFMH over competing schemes is clear, with an mESS/s that is approximately 80 times larger than MJP / PMMH. There is relatively little difference between the performance of LNA / CPMMH and MJP / PMMH.  

Table~\ref{tab:tabEpi1} also includes summarised posterior output when using a deterministic ODE model of latent incidence, combined with a Guassian approximation to the Binomial observation model (ODE / MH). This approach requires only the solution of the ODE system in (\ref{lna1}), as opposed to (\ref{lna1}), (\ref{lna2}) and (\ref{lna3}) when using the LNA. Consequently, an approximate 4-fold increase in overall efficiency is achieved for ODE / MH compared to LNA / FFMH. However, ignoring intrinsic stochasticity leads to a clear loss of inferential accuracy. In particular, the reporting rate is underestimated (which is unsurprising, as this leads to a larger observation variance, which can somewhat offset the inability of the latent ODE model to capture intrinsic stochasticity) and the basic reproduction number is typically overestimated. As $N_{\textrm{pop}}$ increases, posterior uncertainty for all static parameters is underestimated (relative to the gold standard MJP approach) irrespective of each data set.   

Figure~\ref{fig:SSpred} shows within-sample predictive summaries (averaged over parameter uncertainty) for the susceptible and infective states under the LNA. That is, for each of $n_{iters}$ parameter draws from the marginal posterior under the LNA, backward sampling (see Section~\ref{sec:ffbs}) was run to obtain samples $(S_t^{(k)},I_t^{(k)})'$ from the conditional posterior $\pi(x_t|\theta^{(k)},\mathcal{D})$ for $t=10,20,\ldots,80$ and $k=1,2,\ldots,n_{iters}$. We see that although the LNA (unsurprisingly) fails to capture the discrete nature of the infective process (most evident for data set $\mathcal{D}_1$), within-sample predictive draws are generally consistent with the ground truth traces generated by the jump process, although there is some suggestion for data sets $\mathcal{D}_1$ and $\mathcal{D}_3$ that the LNA is least accurate at the end of the epidemic.

\subsection{Oak processionary moth in Richmond park, London}
\label{sec:opm}

In this section we consider the application of the methodology to the infestation of the oak processionary moth (OPM),  \textit{Thaumetopoea processionea}, in Richmond Park, London. OPM is an invasive pest, destructive to oak trees and toxic to humans and animals \citep{Maier03,Maier04,Gottschling06,Rahlenbeck15}. The moth was first established in the UK in 2006 and despite efforts to initially eradicate, and then contain the infestation, OPM has continued to spread \citep{Suprunenko21,Wadkin22}. 

Surveys and control strategies for Richmond Park are carried out by The Royal Parks charity, and this data is then shared with the governmental Oak Processionary Moth Control Programme \citep{OPMForestryCommission}. The data records the numbers of OPM nests removed from trees (with recorded locations) between the years 2013 and 2020, allowing the formation of a time series for the yearly removal incidence of infested trees; see Table~\ref{tab:tabOPM1}. The removal prevalence of the same set of trees (constructed under the assumption of known initial conditions) was considered in an SIR model in \cite{Wadkin22}. However, upon the manual removal of the OPM nests, it is possible for the trees to become susceptible to re-infestation, and thus we additionally consider the SIRS model below.  

\begin{table}[t]
\centering
\small
	\begin{tabular}{@{}l lll lll ll@{}}
         \toprule
Year         & 2013 & 2014 & 2015 & 2016 & 2017 & 2018 & 2019 & 2020 \\   
\midrule
No. removals & 1024 & 1414 & 958 & 540 & 594 & 557 & 587 & 1029 \\
\bottomrule
\end{tabular}
      \caption{OPM data. Number of ``removed trees'' in a given year, Richmond park, London, 2013--2020.}\label{tab:tabOPM1}	
\end{table}

\subsubsection{Model and prior distribution}
\label{sec:opmModels}

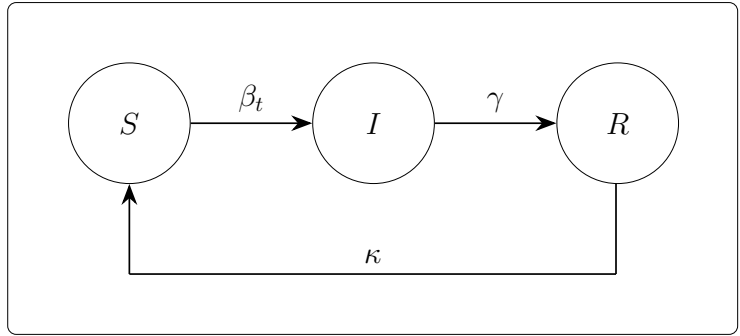
\begin{figure}[ht!]
\centering
\scalebox{0.8}{
{\Large
\begin{tikzpicture}
\draw [rounded corners,fill=white!20] (-2,-3.5) rectangle (10,2);
\node (S) [draw, circle, minimum height=2cm,fill=white] {$S$};
\node (I) [draw, circle, xshift=1cm, minimum height=2cm,right = of S,fill=white] {$I$};
\node (R) [draw, circle, xshift=1cm, minimum height=2cm,right = of I,fill=white] {$R$};
\draw [arrow] (S) -- node[anchor=south] {${\beta}_t$} (I);
\draw [arrow] (I) -- node[anchor=south] {$\gamma$} (R);
\draw [thick] (8,-1) -- (8,-2.5);
\draw [thick] (8,-2.5) -- node[anchor=south] {$\kappa$} (0,-2.5);
\draw [arrow] (0,-2.5) -- (S);
\end{tikzpicture}
}
}
\caption{SIRS compartment model.}\label{fig:SIRS}
\end{figure}

To allow for removed trees re-entering the susceptible class, we consider the SIRS compartment 
model shown graphically in Figure~\ref{fig:SIRS}.
Transitions between compartments can be described by the set of pseudo-reactions given by
\[
S+I \xrightarrow{\beta_t} 2I, \quad I \xrightarrow{\gamma} R, \quad R \xrightarrow{\kappa} S
\]
where $\beta_t$ is a time varying infection rate whose natural logarithm is described by the SDE
\[
d\log\beta_t = \sigma_{\beta}\,dW_t.
\]
That is, the log infection rate is a scaled Brownian motion process. We note that 
setting $\kappa=0$ gives the SIR model and in what follows fit both SIR and SIRS 
models under the assumption the latent incidence process is well described 
by the linear noise approximation. Further details of the LNA for the SIRS 
compartment model are given in Appendix~\ref{sec:appSIRS}. We additionally consider two observation models; these 
are the Binomial and Negative Binomial models given by (\ref{obsBin}) and (\ref{obsNBin}). 
This leads to 4 competing models which we compare using the deviance information 
criterion \citep[DIC, see e.g.][for a discussion of DIC in the epidemic context]{gibson18} 
given by
\[
\textrm{DIC} = -2 \textrm{E}_{\theta}\{\log \pi(y|\theta)|y\} + p_{D}
\]  
where $p_{D}= -2 \textrm{E}_{\theta}\{\log \pi(y|\theta)|y\} + 2\log \pi(y|\bar{\theta})$ measures 
the effective number of parameters in the model. Note that the observed data likelihood 
$\pi(y|\theta)$ is tractable under the Gaussian approximation approach to inference described in 
Section~\ref{sec:ffbs}, which we employ here. Hence, DIC is easily calculated and 
the model with the smallest DIC is preferred.  
 
We follow \cite{Wadkin22} by fixing $N_{\textrm{pop}}=40,000$, $x_0=(38600,1400)'$ and 
$\log\beta_0=-10$. We adopt an independent prior specification by taking 
$\log \gamma\sim \textrm{N}(0,0.5^2)$, $\log \kappa \sim \textrm{N}(0,1)$, $\log\sigma_\beta \sim \textrm{N}(1,1)$, 
$\lambda\sim \textrm{Unif}(0,1)$ and, when using a Negative Binomial observation model, $\log\phi\sim \textrm{N}(0,1)$. 
Note that the choice of $\beta_0$ and prior for the removal rate $\gamma$ induces a 
prior on the basic reproduction number $\mathcal{R}_0=N_{\textrm{pop}}\beta_0 /\gamma$ at time $0$ 
as lognormal $\textrm{logN}(0.6,0.5^2)$. This gives a 95\% equitailed credible interval of $(0.7,4.8)$ 
for $\mathcal{R}_0$, which reflects our belief that OPM spread is likely to persist, without 
precluding $\mathcal{R}_0<1$. We note also that the prior for $\kappa$ gives a $95\%$ credible 
interval of $(0.14,7.10)$ years, reflecting vague prior beliefs on the time taken for a removed 
tree to re-enter the susceptible class.

\subsubsection{Results}
\label{sec:opmResults}

\begin{figure}[ht!]
\centering
\psfragscanon
\psfrag{Ga}[][][0.7][0]{$\gamma$}
\psfrag{Ka}[][][0.7][0]{$\kappa$}
\psfrag{sigB}[][][0.7][0]{$\sigma_{\beta}$}
\psfrag{Lam}[][][0.7][0]{$\lambda$}
\includegraphics[width=5.0cm,height=13cm,angle=-90]{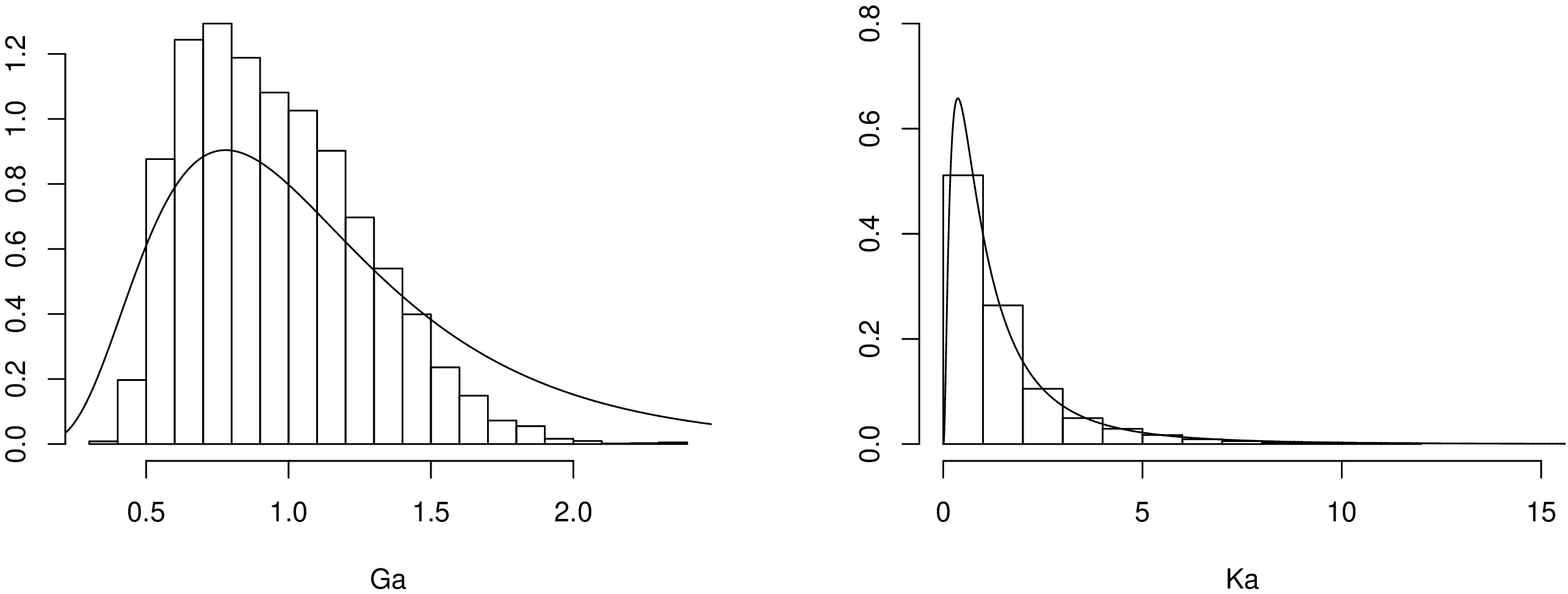}
\includegraphics[width=5.0cm,height=13cm,angle=-90]{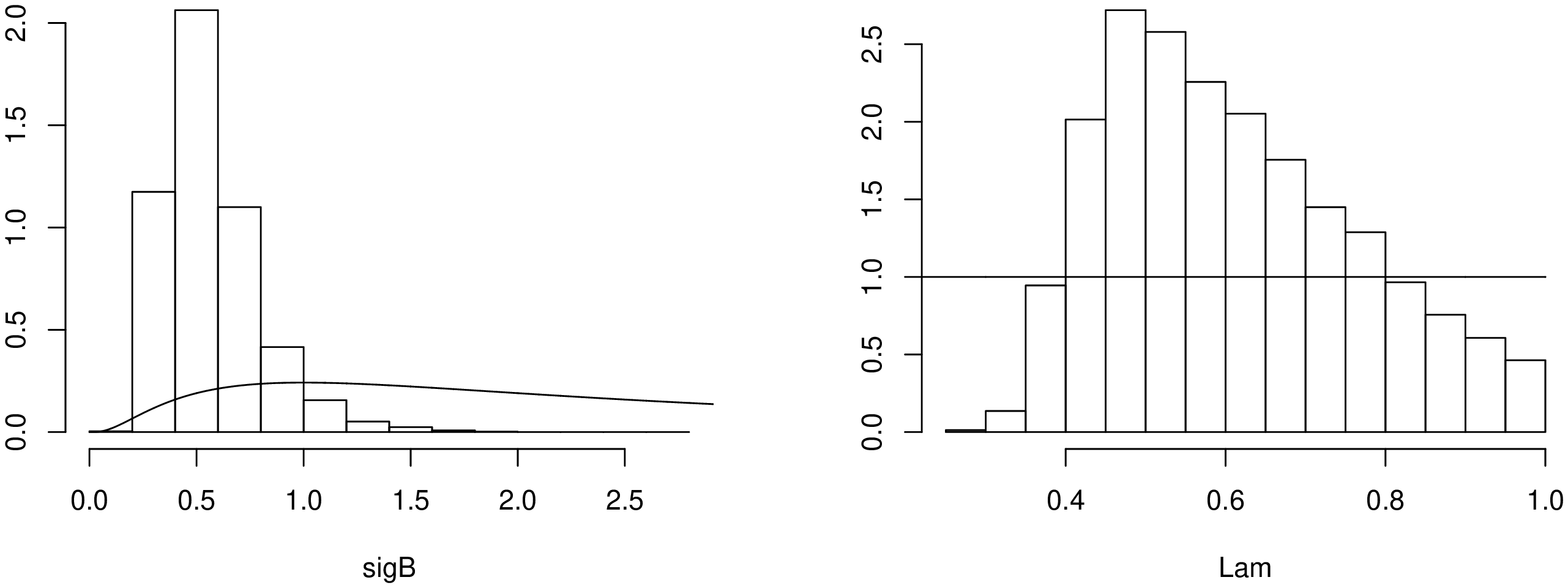}
\caption{OPM data application. Marginal posterior densities (histograms) and prior (solid line), of the parameters in the SIRS model assuming binomial observations.}
\label{fig:OPMdens}
\end{figure}

\begin{figure}[ht!]
\centering
\psfragscanon
\psfrag{year}[][][0.7][0]{year}
\psfrag{St}[][][0.7][-90]{$S_t$}
\psfrag{It}[][][0.7][-90]{$I_t$}
\psfrag{logBt}[][][0.7][-90]{$\log\beta_t$}
\includegraphics[width=5.5cm,height=16cm,angle=-90]{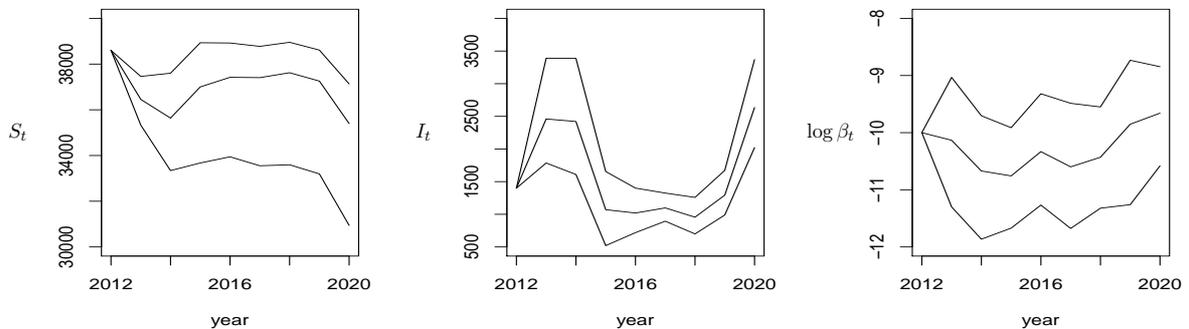}
\caption{OPM data application. Within-sample predictive distributions (mean and 95\% credible intervals) for $S_t$ (left) and $I_t$ (middle) and $\log\beta_t$ (right).}
\label{fig:OPMpred}
\end{figure}

\begin{figure}[ht!]
\centering
\psfragscanon
\psfrag{year}[][][0.7][0]{year}
\psfrag{R0t}[][][0.7][-90]{$\mathcal{R}_{0,t}$}
\includegraphics[width=7cm,height=12cm,angle=-90]{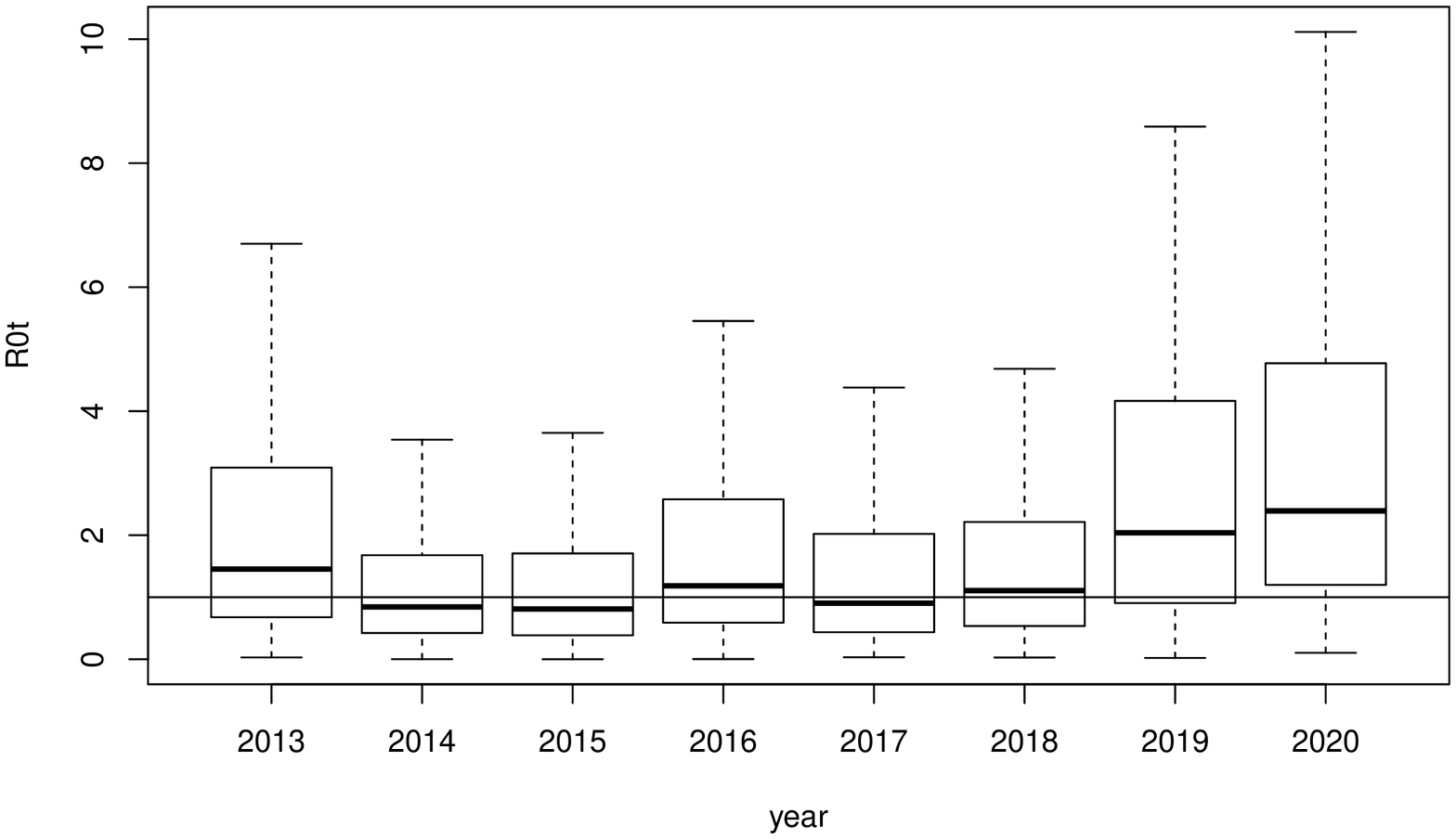}
\caption{OPM data application. Boxplots summarising the marginal posterior distribution of the basic 
reproduction number $\mathcal{R}_0$ against year.}
\label{fig:OPMrep}
\end{figure}

We ran the marginal Metropolis-Hastings scheme (as described in Section~\ref{sec:ffbs}) 
for $50,000$ iterations, with the resulting parameter chains suggesting adequate 
mixing. Tables~\ref{tab:tabOPM2}--\ref{tab:tabOPM3} and Figures~\ref{fig:OPMdens}-\ref{fig:OPMrep} summarise the posterior 
output under each competing model.  

Table~\ref{tab:tabOPM2} shows estimated DIC for the SIR and SIRS models, under the assumption of 
either a Binomial or Negative Binomial observation model. A Binomial observation model is 
preferred irrespective of the assumed underlying compartment model. This is consistent with the 
inferred values of the (inverse) dispersion parameter $\phi$, which are typically small; see Table~\ref{tab:tabOPM3}. 
Hence, it appears that the true but unobserved removal incidence is much larger than the observed incidence, which 
is unsurprising given surveyed areas of Richmond park in each year, which typically consitute a small 
fraction of the total area. Although our findings support the hypothesis that 
trees can become susceptible to re-infestation over the time scales 
of the data set considered, we note that the analysis has not been particularly informative for the parameter 
$\kappa$, governing the rate of $R$ to $S$ transitions; see Figure~\ref{fig:OPMdens} showing marginal 
posterior densities for parameters in the SIRS model and the prior specification. Since removed trees are 
treated with insecticide, this parameter is likely to be of interest to practitioners. Nevertheless, 
improved data collection protocols and a longer study period may provide a partial record of the number of 
$R$ to $I$ transitions, which would greatly improve inferences on $\kappa$.    

Figure~\ref{fig:OPMpred} summarises the within-sample predictive distributions for the susceptible and infective 
prevalence processes (which are easily reconstructed from the predicted incidences, not shown) and the 
log infection process, $\log\beta_t$. These results are broadly consistent with those of \cite{Wadkin22}, 
which suggest a plausibly constant infection rate and an uptick in infected trees from 2018. Figure~\ref{fig:OPMrep} 
summarises the marginal posterior distribution of the basic 
reproduction number $\mathcal{R}_0$ against year. Sampled values of $\mathcal{R}_{0}$ appear to be largely 
consistent across years, however, the marginal posterior distributions in years 2018, 2019 and 2020 have 
greatest support for $\mathcal{R}_0>1$ suggesting that OPM will continue to propagate in Richmond park.   

\begin{table}[t]
\centering
\small
	\begin{tabular}{@{}l lll l@{}}
         \toprule
Model & SIR (Bin) & SIR (Neg Bin) & SIRS (Bin) & SIRS (Neg Bin) \\   
\midrule
DIC   & 115.4  & 120.4 & 113.4 & 119.8 \\
\bottomrule
\end{tabular}
      \caption{OPM data application. Estimated DIC for the SIR and SIRS models, assuming either a Binomial (Bin) 
or Negative Binomial (Neg Bin) observation model.}\label{tab:tabOPM2}	
\end{table}

\begin{table}[t]
\centering
\small
	\begin{tabular}{@{}l lll l@{}}
         \toprule
 & \multicolumn{4}{c}{Mean (Standard Deviation)}\\
\cmidrule(l){2-5}
Parameter  & SIR (Bin) & SIR (Neg Bin) & SIRS (Bin) & SIRS (Neg Bin)  \\   
\midrule
$\gamma$       & 0.90 (0.30) & 1.44 (0.44) & 0.96 (0.31) & 1.45 (0.47) \\
$\kappa$       & \phantom{0.00} -- \phantom{0.00} & \phantom{0.00} -- \phantom{0.00} & 1.57 (2.02) & 1.26 (1.72) \\
$\sigma_\beta$ & 0.64 (0.24) & 0.58 (0.58) & 0.56 (0.22) & 0.51 (0.39) \\
$\lambda$      & 0.64 (0.16) & 0.57 (0.20) & 0.61 (0.16) & 0.62 (0.20)\\
$\phi$         & \phantom{0.00} -- \phantom{0.00} & 0.30 (0.38) & \phantom{0.00} -- \phantom{0.00}  & 0.25 (0.32) \\
\bottomrule
\end{tabular}
      \caption{OPM data application. Marginal parameter posterior summaries.}\label{tab:tabOPM3}	
\end{table}

\section{Discussion}
\label{sec:sum}

The construction of efficient and widely applicable approaches 
to inference for stochastic epidemic models remains a key challenge 
\citep{Swallow22}. In this paper, we have proposed a fast an efficient method 
for inferring the parameters governing the linear noise approximation (LNA) 
of a stochastic epidemic model, using incidence data consisting of 
the cumulative number of new infections (or removals) in fixed-length 
windows. This setting is considered in \cite{fintzi2021linear} who combine 
the LNA of the incidence process with a Negative Binomial observation model 
and develop an efficient MCMC scheme targeting the joint density of the 
parameters and latent incidence process. Our contribution, on the 
other hand, is a framework for marginalising out the latent incidence 
process: either by exactly targeting the marginal parameter posterior 
via a (correlated) pseudo-marginal method, or analytically through a Gaussian 
approximation of the observation process. We additionally allow for a 
flexible, time-varying stochastic infection rate, which is naturally 
handled within the LNA framework. Our experiments demonstrated 
that use of the LNA and a further Gaussian approximation of an observation 
model can be both accurate and efficient. Using parameter values inspired 
by the Abakaliki smallpox outbreak, we investigated the accuracy and efficiency 
of the analytically marginalised LNA as the population size increases 
(and with the parameters scaled appropriately). In the `large epidemic' 
setting ($N_{\textrm{pop}}=1200$), the analytic marginalisation scheme 
outperforming the next best performing scheme by about a factor of 80. In this 
scenario, use of the most natural Markov jump process representation of 
the epidemic is computationally prohibitive.

We further illustrated our approach via an application with real data 
consisting of numbers of trees infested with oak processionary moth (OPM) nests 
in Richmond Park, London. Typical observations consist of around 500--1500 
removals in a given year, with a total population size of around $40,000$ 
trees, thus necessitating the efficient inference 
methods developed here. As well as inferring key quantities of interest, 
such as the basic reproduction number and latent susceptible and infective 
trajectories, our approach allows for easy computation of the observed data 
likelihood which can be used, for example, to compute a deviance information 
criterion (DIC). We used DIC to compare two different compartment models 
(SIR versus SIRS) and two different observation models (Binomial versus 
Negative Binomial). Out analysis suggests the SIRS model as the best fitting 
compartmentment model, suggesting that trees can re-enter the susceptible 
class following removal (via treatment). Although improved data collection 
protocols which include observation of the number of removed trees which 
susbequently become infected will greatly improve predictive power, our approach 
demonstrates that meaningful conclusions on the spread of OPM can be drawn, despite 
a data poor scenario.      

\subsection{Limitations and extensions} 
\label{sec:lim}
Within the stochastic kinetic models context, the LNA can be derived 
directly from the most natural Markov jump process (MJP) representation \citep[][]{kurtz1970,kurtz1972} 
but is perhaps most intuitively viewed as a tractable Gaussian process approximation 
of the It\^o Stochastic differential equation (SDE) 
that best matches the MJP representation \cite[][]{ferm2008,fearnhead14}. As 
advocated by \cite{Fuchs_2013} among others, judging the validity of these 
continuous-valued approximations should involve comparison with the MJP 
(e.g. via simulations) for the specific system considered. Nevertheless, 
we expect, in general, that the best matching SDE and LNA approaches are likely 
to approximate the MJP particularly poorly when specie numbers are comparatively small 
(e.g. in the few tens). In such situations, we envisage that our approach is likely 
to be of most practical benefit in providing initial values and tuning choices for 
simulation-based inference schemes that target the posterior under the MJP. For inherently 
multi-scale epidemics, it may be possible to leverage hybrid simulation techniques 
\citep[see e.g.][]{sherlock2014} whereby the LNA is used to model species which 
frequently change state, coupled with a discrete stochastic updating procedure for 
species which change state less often. 

This work can be further extended in a number of ways. Of particular interest to us is the 
use of the proposed approach within a spatio-temporal setting, and with application 
to OPM spread, for example, by allowing importation of pests from nearby locations. 
Extension of the methodology to allow incorporation of multiple data streams 
\citep[see e.g.][]{corbella22} also merits further attention.

\subsection*{Acknowledgements}

We thank Julia Branson from the University of Southampton and Andrew Hoppit 
from Forestry Commission England for useful discussions regarding the OPM 
application. This research was supported by: EPSRC New Horizons Grant EP/V048511/1 
(AB, AG, NP, and LW) and NERC Knowledge Exchange Fellows Grant NE/X000478/1 (LW).

\appendix

\section{SIRS model details}
\label{sec:appSIRS}
Let $X_t=(S_t,I_t)'$ denote the numbers of susceptibles and infectives at time $t$. 
Similarly, let $n_t=(n_{1,t},n_{2,t},n_{3,t})'$ denote the cummulative number of 
infection, removal and loss of immunity (that is, removal to susceptible) events 
at time $t$. Let $\beta_t$, $\gamma$ and $\kappa$ denote the corresponding event rates. 
The cumulative incidence $\{N_t, t\geq 0\}$ is an MJP governed by the transition probabilities 
\begin{align*}
\mathbb{P}(N_{t+\Delta t}=(n_{1,t}+1,n_{2,t},n_{3,t})'|n_t,x_t,\theta) &= \beta_t s_t i_t \,\Delta t+o(\Delta t),\\
\mathbb{P}(N_{t+\Delta t}=(n_{1,t},n_{2,t}+1,n_{3,t})'|n_t,x_t,\theta) &= \gamma i_t \,\Delta t+o(\Delta t),\\
\mathbb{P}(N_{t+\Delta t}=(n_{1,t},n_{2,t},n_{3,t}+1)'|n_t,x_t,\theta) &= \kappa (N_{pop}-s_t-i_t) \,\Delta t+o(\Delta t),\\
\mathbb{P}(N_{t+\Delta t}=(n_{1,t},n_{2,t},n_{3,t})'|n_t,x_t,\theta) &= 1-(\beta s_t i_t + \gamma i_t+\kappa [N_{pop}-s_t-i_t])\,\Delta t+o(\Delta t)
\end{align*}
and recall that $N_{pop}$ is the total population size (assumed fixed and known). The stoichiometry matrix is given by 
\[
S=\begin{pmatrix}
 -1 & 0  & 1  \\
  1 & -1 & 0
\end{pmatrix}
\]
and the hazard function is 
\[
h(x_t)= (\beta_t s_t i_t, \gamma i_t, \kappa [N_{pop}-s_t-i_t])'.
\]
Using (\ref{MJP1}), we may write the hazard function in terms of the incidence process as 
\[
h^*(n_t)=(\beta [s_0-n_{t,1}+n_{t,3}][i_0+n_{t,1}-n_{t,2}],\gamma [i_0+n_{t,1}-n_{t,2}],\kappa[N_{pop}-s_0-i_0-n_{t,3}+n_{t,2}])'.
\] 
Now define $N_{4,t}=\log \beta_t$ as a Brownian motion process scaled by $\sigma_{\beta}$. 
This leads to the CLE for the SIRS model (with time varying infection rate) as 
\[
dN_{t} = \left\{h_{1}^*(n_t),h_{2}^*(n_t),h_{3}^*(n_t),0 \right\} dt + 
\operatorname{diag}\left\{\sqrt{h^*_1(n_t)},\sqrt{h^*_2(n_t)},\sqrt{h^*_3(n_t)},\sigma_{\beta}\right\}\, dW_t
\]
where $W_t$ is a length-$4$ vector of uncorrelated Brownian motion processes. The LNA then follows from equations 
(\ref{lna1}), (\ref{lna2}) and (\ref{lna3}), with the (transpose of the) Jacobian matrix $F_t$ given by
\[
F_t'=\begin{pmatrix}
\exp(\eta_{4,t}) (s_0-i_0-2\eta_{t,1}+\eta_{t,2}+\eta_{3,t}) & \gamma & 0 & 0\\
\exp(\eta_{4,t}) (\eta_{t,1}-\eta_{t,3}-s_0) & -\gamma & \kappa & 0\\ 
\exp(\eta_{4,t}) (i_0+n_{t,1}-n_{t,2}) & 0 & -\kappa & 0 \\
\exp(\eta_{4,t})(s_0-\eta_{t,1}+\eta_{t,3})(i_0+\eta_{t,1}-\eta_{t,2}) & 0 & 0 & 0
\end{pmatrix}.
\]

\bibliographystyle{apalike}
\bibliography{references}

\end{document}